\documentclass[amsmath,amssymb,showkeys,nofootinbib,superscriptaddress,aps,prd,10pt,twocolumn]{revtex4-2}
\usepackage{graphicx}
\usepackage[utf8]{inputenc}
\usepackage[caption=false]{subfig}
\usepackage{multirow}
\usepackage{appendix}
\usepackage{graphicx,epstopdf}

\usepackage{colortbl}
\usepackage{xcolor}
\usepackage[colorlinks=true, urlcolor=blue, linkcolor=blue, citecolor=blue]{hyperref}
\usepackage{float}

\begin{document}

\title{The $T_{bc}$ tetraquarks near the $B\bar{D}$ threshold}
\author{Halil Mutuk}%
\email[]{hmutuk@omu.edu.tr}
\affiliation{Department of Physics, Faculty of Sciences, Ondokuz Mayis University, 55139 Samsun, Türkiye}

 
\begin{abstract}
We study the doubly heavy open-flavor tetraquarks $T_{bc}^{(0)}$ ($J^{P}=0^{+}$) and $T_{bc}^{(1)}$ ($J^{P}=1^{+}$) in the dynamical diquark model, describing the system as a heavy antidiquark--light diquark pair interacting through the lattice-QCD $\Sigma_g^+(1S)$ Born--Oppenheimer potential. Solving the radial Schr\"odinger equation yields
$M(T_{bc}^{(0)}) = 7.143$--$7.158$ GeV and
$M(T_{bc}^{(1)}) = 7.217$--$7.222$ GeV, with hyperfine splittings of $\Delta_{HF}\simeq 59$--$79$ MeV. The splitting is driven mainly by the mass difference between symmetric and antisymmetric heavy-antidiquark configurations, while the chromomagnetic interaction contributes linearly with $\partial\Delta_{HF}/\partial\kappa_{\bar b\bar c}=2$, consistent with heavy-antidiquark spin algebra. The mean separation,
$\langle r\rangle\simeq 0.45$--$0.46$ fm, and inverse radius,
$\langle 1/r\rangle^{-1}\simeq 0.33$--$0.34$ fm, exhibit weak parameter dependence and are consistent with a compact diquark--antidiquark interpretation. Relative to open-flavor thresholds, the central value of the scalar mass lies essentially at the $B\bar D$ threshold, suggesting that this state may appear either as a weakly decaying bound tetraquark or as a narrow near-threshold structure. The axial-vector eigenvalue, in turn, consistently lies $23$--$28$ MeV above $B^{*}\bar D$ and about $70$ MeV below $B\bar D^{*}$, indicating a natural candidate for an $S$-wave resonance whose line shape would be strongly influenced by the nearby $B^{*}\bar D$ threshold. We quantify the parametric uncertainties of these estimates, which are dominated by the light-diquark mass and largely cancel in mass differences, and discuss the limitations inherent in identifying single-channel eigenvalues with physical pole positions in the near-threshold regime. We further complete the $\Sigma_g^+(1S)$ multiplet, showing that its four isovector members are obtained from the isoscalar ones by a rigid shift of $2\kappa_{ud}\approx 206$ MeV and lie above every $B^{(*)}\bar D^{(*)}$ threshold, and we estimate the effect of the neglected $\bar{\mathbf{6}}\otimes\mathbf{6}$ color configuration, which despite the attractive $\mathbf{6}$--$\bar{\mathbf{6}}$ inter-cluster potential lies $0.1$--$0.3$ GeV higher and contributes only at second order.
\end{abstract}

\maketitle

\section{Introduction}\label{intro}

The past two decades have witnessed a dramatic expansion of the known hadron spectrum, with the discovery of numerous exotic states that cannot be accommodated within the conventional $q\bar{q}$ and $qqq$ quark model. Among these, manifestly flavor-exotic multiquark states occupy a particularly important position, since they cannot mix with ordinary mesons or baryons and therefore provide an especially clean probe of nonperturbative QCD dynamics. The recent observation by LHCb of the doubly-charmed tetraquark $T_{cc}^+$~\cite{LHCb:2021vvq,LHCb:2021auc}, with a mass only about 360~keV below the $D^0 D^{*+}$ threshold, has definitively established that antiheavy-antiheavy-light-light systems $\bar{Q}\bar{Q}'qq'$ can form near-threshold bound structures in nature, and has sharpened interest in their bottom-sector analogues.

From a theoretical standpoint, $\bar{Q}\bar{Q}'qq'$ tetraquarks are among the most promising candidates for strong-interaction-stable exotic hadrons. The possibility that four-quark configurations containing heavy quarks may be stable against strong decay was first raised in the pioneering study of Ref.~\cite{Ader:1981db}, which predates all subsequent model analyses of doubly-heavy systems. In the heavy-quark limit $m_Q \to \infty$, the attractive color-Coulomb interaction between two heavy antiquarks in a color-$\mathbf{\bar{3}}_c$ configuration guarantees the existence of bound states~\cite{Ader:1981db,Carlson:1987hh,Manohar:1992nd,Eichten:2017ffp}, in close analogy with the role played by a heavy quark in singly-heavy baryons. The binding is reinforced by the attraction of the two light quarks in Jaffe's ``good diquark'' configuration ($\mathbf{\bar{3}}_c$, spin $0$)~\cite{Jaffe:2004ph}, and both mechanisms are absent, or strongly suppressed, in a configuration of two well-separated heavy-light mesons. These considerations single out the $I(J^P)=0(1^+)$ $ud\bar{b}\bar{b}$ and $I(J^P)=\tfrac{1}{2}(1^+)$ $\ell s\bar{b}\bar{b}$ channels as the most favorable candidates for binding, and indeed multiple independent lattice QCD calculations have now firmly established the existence of deeply bound, strong-interaction-stable tetraquarks in both~\cite{Francis:2016hui,Junnarkar:2018twb,Leskovec:2019ioa,Hudspith:2020tdf,Meinel:2022lzo}.

The doubly-bottom tetraquarks, however, are experimentally challenging because their production requires two $b\bar{b}$ pairs. The mixed bottom-charm system $T_{bc}$ is considerably more attractive in this regard: simultaneous $b\bar{b}$ and $c\bar{c}$ production is already well established through the copious observation of $B_c$ mesons, and the estimated production cross section of a $T_{bc}$ tetraquark at the LHC is about a factor of forty larger than that of its doubly-bottom counterpart~\cite{Ali:2018xfq}. Whether a bound state actually exists in this channel is, however, a much more delicate question than in the doubly-bottom case. Replacing one $\bar{b}$ by $\bar{c}$ reduces the reduced mass of the heavy antidiquark, and hence the strength of the short-distance color-Coulomb attraction, and simultaneously enhances the residual spin-dependent heavy-light interactions that are known from the heavy-baryon spectrum to oppose binding. The $T_{bc}$ system thus sits at the borderline between bound and unbound, and its study provides a stringent test of our understanding of the dynamics responsible for exotic binding.

The theoretical situation for $T_{bc}$ has, until very recently, been strikingly inconclusive. Heavy-quark symmetry arguments~\cite{Eichten:2017ffp} predict no bound state in either the $I(J^P)=0(0^+)$ or $0(1^+)$ channel. Most nonrelativistic and relativized quark models~\cite{SilvestreBrac:1993ss,Semay:1994ht,Ebert:2007rn,Park:2018wjk,Carames:2018tpe,Braaten:2020nwp,Lu:2020rog} find at most very weakly bound or unbound ground states, while many chiral quark model analyses~\cite{Deng:2018kly,Yang:2019itm,Tan:2020ldi} predict substantial binding of order $130$--$220$~MeV. QCD sum-rule studies are similarly scattered, ranging from unbound or modestly bound to very deeply bound~\cite{Chen:2013aba,Agaev:2018khe}. A coherent picture has therefore had to come from first-principles lattice QCD.

Direct lattice QCD studies of the $T_{bc}$ system have themselves evolved considerably, and initially produced apparently contradictory results. The first such calculation employed nonrelativistic QCD (NRQCD) for the bottom quark and a relativistic heavy-quark action for the charm quark on three PACS-CS ensembles with pion masses reaching $m_\pi \simeq 164$~MeV \cite{Francis:2018jyb}. That work reported evidence for a strong-interaction-stable $I(J^P)=0(1^+)$ ground state with a binding energy in the range $-61~\mathrm{MeV} < \Delta E < -15~\mathrm{MeV}$ relative to the $D\bar{B}^*$ threshold, placing the predicted mass near the electromagnetic $D\bar{B}\gamma$ decay threshold. The lightest ensemble, however, had $m_\pi L = 2.4$, raising concerns about finite-volume effects, and the plateaus of the wall-local correlators were short and late-time. A follow-up study \cite{Hudspith:2020tdf}, on a larger-volume ensemble with $m_\pi \simeq 192$~MeV and $m_\pi L = 4.2$, and employing an improved ``box-sink'' construction together with an expanded operator basis, did not confirm the earlier signal: the $I(J^P)=0(1^+)$ $T_{bc}$ ground state was found to lie marginally above the $D\bar{B}^*$ threshold, with no evidence for binding in either the $0^+$ or $1^+$ channel. The earlier indication of binding was attributed to late-onset wall-local plateaus and residual finite-volume effects. Subsequently, Ref.~\cite{Meinel:2022lzo}, using domain-wall light quarks, NRQCD bottom quarks, and an anisotropic clover action for the charm, and crucially including nonlocal meson-meson (scattering) interpolating operators at the sink on five ensembles including one at the physical pion mass, studied both the $0(0^+)$ and $0(1^+)$ channels. They found finite-volume ground-state energies consistent with the respective lowest two-meson thresholds $BD$ and $BD^*$, and concluded that there was no clear evidence of binding in either channel, although a shallow bound state close to threshold could not be ruled out.

Two very recent lattice QCD studies have substantially clarified this picture and, for the first time, provided consistent evidence for binding in the $T_{bc}$ system. Ref.~\cite{Padmanath:2023rdu} performed a calculation on four gauge-field ensembles generated by the MILC Collaboration with $N_f = 2+1+1$ dynamical quark flavors, employing the Highly Improved Staggered Quark (HISQ) action. The ensembles span lattice spacings from $\sim 0.058$ to $\sim 0.12$~fm and include five valence light-quark masses, with a correlation-matrix basis constructed from both local diquark--antidiquark operators and local two-meson $D\bar{B}^*$ and $B\bar{D}^*$ interpolators. Extracting the $D\bar{B}^*$ $S$-wave scattering length via L\"uscher's method and performing a continuum and chiral extrapolation, they found a positive scattering length $a_0^{\mathrm{phys}} = +0.57(^{+4}_{-5})(17)$~fm at the physical pion mass, corresponding to a genuine $I(J^P)=0(1^+)$ $T_{bc}$ bound state $T_{bc}$ with a binding energy of $-43(^{+6}_{-7})(^{+14}_{-24})$~MeV relative to the $D\bar{B}^*$ threshold. The authors further observed that the strength of the binding decreases with increasing $m_{u/d}$, and that the system becomes unbound above a critical pseudoscalar mass $M_{\mathrm{ps}}^* = 2.73(21)(19)$~GeV. Complementing this work, Ref.~\cite{Alexandrou:2023cqg} carried out the first lattice-QCD determination of the full energy dependence of the $B\text{-}\bar{D}$ and $B^*\text{-}\bar{D}$ isospin-$0$, $S$-wave scattering amplitudes both below and above threshold, using large $7\times 7$ and $8\times 8$ symmetric correlation matrices with $B^{(*)}\text{-}\bar{D}$ scattering operators at both source and sink and with nonzero back-to-back momenta. Working at a single lattice spacing and pion mass $m_\pi \simeq 220$~MeV on two volumes, and fitting the extracted phase shifts with effective-range expansions, they found near-threshold poles in both the $J=0$ and $J=1$ channels corresponding to shallow bound states---either genuine or virtual---only a few MeV below the respective $B^{(*)}\text{-}\bar{D}$ thresholds, together with hints of broad resonances of width $\sim 200$~MeV located about $100$~MeV above threshold. Taken together, these two studies---while differing in the depth of the predicted binding and in the resolution of the genuine-versus-virtual bound-state question---both point toward attractive $B^{(*)}\bar{D}$ interactions strong enough to produce near-threshold poles, and mark a qualitative shift in the theoretical status of the $T_{bc}$ system.

Despite this recent progress, several important questions remain open. The binding energy of the $I(J^P)=0(1^+)$ $T_{bc}$ is quoted as $-43(^{+6}_{-7})(^{+14}_{-24})$~MeV in Ref.~\cite{Padmanath:2023rdu} but at most a few MeV in Ref.~\cite{Alexandrou:2023cqg}, and the genuine versus virtual character of the near-threshold poles, the existence of the $I(J^P)=0(0^+)$ partner, and the nature of the possible broad resonances above threshold all require further scrutiny. Left-hand cut effects from pion exchange, coupled-channel dynamics involving $BD$, $B^*D$, $BD^*$, and $B^*D^*$, discretization effects, and the chiral extrapolation all contribute to the current systematic uncertainties and motivate additional independent calculations with different lattice actions, ensembles, and analysis strategies. 

In the present work, we compute the masses, hyperfine splittings, and mean inter-cluster radii of the doubly-heavy open-flavor tetraquarks
$T_{bc}^{(0)}$ ($J^{P}=0^{+}$) and $T_{bc}^{(1)}$ ($J^{P}=1^{+}$) within the Born--Oppenheimer dynamical diquark model, using the
$\Sigma_g^+(1S)$ potential extracted from lattice QCD. Comparing the resulting spectrum with the
lowest $B^{(*)}\bar{D}^{(*)}$ thresholds, we find that the central value of the scalar mass straddles the $B\bar{D}$ threshold to within a few MeV---thus
sitting on the boundary between a genuinely bound and a narrow near-threshold tetraquark---while the axial-vector eigenvalue consistently lies $\sim 25$~MeV above $B^{*}\bar{D}$, making this state a natural candidate for an $S$-wave resonance with discovery prospects at LHCb and Belle~II. Since the states of interest sit in the immediate vicinity of open two-meson thresholds, we devote particular attention to the systematic limitations of the single-channel constituent description employed here: the restriction to the $\bar{\mathbf{3}}_c\otimes\mathbf{3}_c$ color configuration, the size of non-adiabatic corrections to the BO approximation, the expected impact of the neglected coupled-channel $B^{(*)}\bar{D}^{(*)}$ dynamics, and the parametric uncertainties of the model inputs, and we formulate our conclusions accordingly.

The remainder of this paper is organized as follows. In Sec.~\ref{sec:method} we summarize the dynamical diquark model in its
Born--Oppenheimer formulation---recalling the role of the $\Sigma_g^+(1S)$ potential and the construction of the heavy-antidiquark--light-diquark Hamiltonian---and describe the numerical solution of the radial Schr\"odinger equation, specifying the input parameters. Section \ref{sec:results} presents our results for the masses, hyperfine splittings, and spatial observables of the $T_{bc}^{(0,1)}$ states, together with a graphical comparison to the relevant $B^{(*)}\bar{D}^{(*)}$ thresholds and a discussion of the resulting phenomenology, including the prospects for experimental discovery. We summarize our findings and outline directions for future work in Sec.~\ref{sec:conclusions}.

\section{Methodology}
\label{sec:method}

\subsection{Born-Oppenheimer approximation}
\label{sec:bo}

The Born-Oppenheimer (BO) approximation was developed in the early days of quantum mechanics~\cite{Born:1927boa} as a tool for studying molecular systems, and exploits the large separation of time scales between the slow motion of heavy atomic nuclei and the fast response of the surrounding electrons. Because the nuclei are much heavier than the electrons, they can be treated as approximately static sources of the electromagnetic field at each instant; the electrons adjust almost instantaneously to the nuclear configuration, and their energy---together with the mutual Coulomb repulsion of the nuclei---defines an effective potential whose eigenvalues determine the vibrational and rotational spectrum of the molecule~\cite{Braaten:2014qka}.

An entirely analogous separation of scales arises in QCD whenever a hadronic system contains heavy color sources. The BO approximation for heavy quarkonium $(Q\bar{Q})$ was first formulated in Ref.~\cite{Juge:1999ie}, where the role of the electrons is played by the gluon field and by any light quarks present in the system, and the role of the nuclei is played by the heavy quark and antiquark. The large ratio $m_Q/\Lambda_{\rm QCD}$ ensures that the gluonic and light-quark degrees of freedom adjust adiabatically to the slowly varying heavy-quark configuration, generating a set of static BO potentials $V_\Gamma(r)$ labelled by the quantum numbers $\Gamma$ of the gluon field. The heavy-quark dynamics is then obtained by solving the Schr\"odinger equation in these potentials, in direct analogy with the molecular case. This picture has since been formulated as a systematic effective field theory (EFT), the Born-Oppenheimer EFT~\cite{Brambilla:2017uyf,Berwein:2024ztx}, and applied to a wide range of exotic hadrons including the $\chi_{c1}(3872)$ and $T_{cc}^+(3875)$~\cite{Brambilla:2024imu}, as well as hidden-charm pentaquarks~\cite{Brambilla:2025xma}. Systematic studies of tetraquarks in the BO framework and in the large-$N$ expansion can be found in Ref.~\cite{Allaman:2024vwn}.

When the BO approximation is applied to four-quark systems, one needs at least two sufficiently heavy color sources, together with degrees of freedom corresponding to light quarks and gluons. The slow sources---here the heavy antiquarks $\bar{b}$ and $\bar{c}$, or, equivalently, the composite clusters built from them---move adiabatically in a confining potential generated by the fast gluonic and light-quark fields~\cite{Maiani:2022qze}. Because the typical excitation energy of the gluon field and the light-quark cloud is of order $\Lambda_{\rm QCD}\simeq 200$--$300$~MeV, as confirmed by lattice-QCD studies of the BO potentials~\cite{Juge:2002br}, while the relative motion of the heavy clusters proceeds on a parametrically slower time scale, the BO expansion provides a controlled framework for the computation of the tetraquark spectrum. The dominant systematic uncertainty of the approach is generically of order $\mathcal{O}(\Lambda_{\rm QCD})$ and affects primarily the fine-structure splittings, while the gross features of the spectrum and the spatial properties of the states are expected to be robust.

For the $T_{bc}$ system considered in this work, both heavy antiquarks are unambiguously ``slow'' in the BO sense, with $m_c,m_b\gg\Lambda_{\rm QCD}$, so that the adiabatic approximation is on an even firmer footing than in the charm-strange case studied recently in Ref.~\cite{Mutuk:2025hql}. The light-quark pair $(u,d)$ and the gluon field are treated as the fast degrees of freedom, whose response to the slowly varying positions of the heavy antiquarks is encoded in an effective BO potential between the color clusters that carry those heavy antiquarks.

The size of the non-adiabatic corrections neglected in this treatment can be estimated as follows. The leading corrections arise from the mixing of the $\Sigma_g^+$ potential with excited BO potentials, induced by the kinetic operator of the heavy clusters, and are parametrically suppressed by the ratio of the characteristic energy of the slow degrees of freedom---the level spacing of the radial spectrum in the $\Sigma_g^+(1S)$ potential, of order a few hundred MeV for the reduced masses relevant here---to the gluonic excitation gap separating $\Sigma_g^+$ from the first excited potentials ($\Pi_u$, $\Sigma_u^-$), which lattice QCD places at roughly $1$~GeV at the relevant inter-cluster separations~\cite{Juge:1999ie,Juge:2002br}. Since the mixing enters the energies at second order, the resulting shifts are expected at the level of a few percent of the radial level spacing, i.e.\ of order $10$~MeV; corrections of this size affect fine-structure quantities but not the gross features of the spectrum. A physically distinct---and, in the present context, more important---limitation is that the adiabatic $\Sigma_g^+$ potential does not describe the rearrangement of the light quarks into color-singlet heavy-light mesons, i.e.\ the coupling of the $\delta$--$\bar\delta'$ configuration to the open $B^{(*)}\bar{D}^{(*)}$ continua. For eigenvalues located within a few tens of MeV of these thresholds, as found below, this coupling can shift the energies by an amount comparable to their displacement from threshold; we return to this point quantitatively in Sec.~\ref{sec:results}. Statements about the near-threshold character of individual states must therefore be understood within these combined $\mathcal{O}(10~\text{MeV})$ systematic limitations of the single-channel adiabatic description, which are addressed explicitly below.

This picture naturally matches onto the dynamical diquark model, which we now describe.

\subsection{Dynamical diquark model}
\label{sec:ddm}

A compact tetraquark can be described as a bound state of a colored diquark and a colored antidiquark, in which the diquark (antidiquark) is treated as an effective degree of freedom. A diquark is a colored bound quark-quark pair $(qq)$ and an antidiquark is a colored bound antiquark-antiquark pair $(\bar{q}\bar{q})$, which can be combined into a color singlet. More detailed discussions of diquark degrees of freedom can be found in the reviews of Refs.~\cite{Anselmino:1992vg,Barabanov:2020jvn}; diquark--antidiquark potential models of this type have been applied broadly across the tetraquark landscape, from all-light~\cite{Lodha:2024fbs} and singly-charmed~\cite{Lodha:2025ijm} systems to doubly-heavy configurations, and unified quark--diquark effective-mass formalisms calibrated on the baryon spectrum have recently been developed for heavy multiquarks~\cite{Mohan:2026blk}. Since a quark transforms in the fundamental representation $\mathbf{3}$ and an antiquark in $\bar{\mathbf{3}}$, combining two quarks gives $\mathbf{3}\otimes\mathbf{3}=\bar{\mathbf{3}}\oplus\mathbf{6}$, while combining two antiquarks gives $\bar{\mathbf{3}}\otimes\bar{\mathbf{3}}=\mathbf{3}\oplus\bar{\mathbf{6}}$. A color-singlet tetraquark $|qq\bar{q}\bar{q}\rangle$ can therefore be formed either from the $\bar{\mathbf{3}}\otimes\mathbf{3}$ or from the $\bar{\mathbf{6}}\otimes\mathbf{6}$ channels.

In the present work we restrict ourselves to the $\bar{\mathbf{3}}\otimes\mathbf{3}$ configuration, and it is important to state explicitly the motivation for, and the limitations of, this choice. At short distances, one-gluon exchange between two quarks is attractive in the $\bar{\mathbf{3}}_c$ channel and repulsive in the $\mathbf{6}_c$ channel, so that only the antitriplet supports compact, quasi-bound diquark clusters. It must be stressed, however, that this argument concerns the interaction \emph{inside} each cluster and does not by itself disfavor the sextet configuration. A $\mathbf{6}_c$ diquark and a $\bar{\mathbf{6}}_c$ antidiquark can be coupled to a color singlet, and the static potential \emph{between} them is attractive; indeed, Casimir scaling---verified on the lattice at the few-percent level for source separations up to $1$~fm~\cite{Bali:2000un}---makes it stronger than the $\bar{\mathbf{3}}_c$--$\mathbf{3}_c$ potential by the factor $C_6/C_3 = 5/2$, so that both the color-Coulomb well is deeper and the confining rise is steeper by the same factor. The relative position of the two color channels is therefore decided by a genuine competition: the repulsive intra-cluster interaction, which prevents the formation of a compact $\mathbf{6}_c$ cluster and thereby raises the configuration in energy, against the enhanced inter-cluster attraction, which lowers it at short distances but penalizes it at the separations $\langle r\rangle\simeq 0.45$~fm relevant here. We examine this competition quantitatively in Sec.~\ref{sec:other} and find that, within the present framework, the net effect places the sextet configuration several hundred MeV above the antitriplet one, so that its admixture acts as a second-order correction rather than as a competing leading configuration. In addition, the dynamical diquark model is built on a specific production mechanism, in which a compact $\bar{\mathbf{3}}_c$ diquark and $\mathbf{3}_c$ antidiquark are formed essentially at a point and separate rapidly before hadronization~\cite{Brodsky:2014xia}; the lattice $\Sigma_g^+$ potential employed below is precisely the static energy of such a color source--antisource pair. The $\bar{\mathbf{3}}\otimes\mathbf{3}$ restriction is therefore internally consistent with the dynamical assumptions of the model rather than an independent approximation. Nevertheless, it must be acknowledged that in a general four-quark Hamiltonian the $\bar{\mathbf{3}}\otimes\mathbf{3}$ and $\bar{\mathbf{6}}\otimes\mathbf{6}$ channels mix, and static considerations alone do not determine the size of the induced sextet admixture. Coupled-color-channel quark-model calculations indicate that this admixture can be non-negligible precisely in doubly-heavy systems in which the heavy pair is not constrained by the Pauli principle: in particular, the $bc\bar{q}\bar{q}$ analysis of Ref.~\cite{Carames:2018tpe}, employing a Bhaduri-type potential and retaining the full color space, finds that the physical eigenstates contain sizable components beyond a single $\bar{\mathbf{3}}\otimes\mathbf{3}$ configuration and that binding conclusions are sensitive to the color basis retained; similar observations have been made in chiral quark-model studies~\cite{Deng:2018kly,Yang:2019itm,Tan:2020ldi}. Since enlarging the variational space can only lower the ground-state energy, the eigenvalues obtained below within the pure $\bar{\mathbf{3}}\otimes\mathbf{3}$ sector should be regarded as upper limits within the model, with the neglected color mixing acting in the direction of additional binding. A quantitative assessment of this effect requires a genuine four-body calculation and lies outside the Born--Oppenheimer diquark framework adopted here; a semiquantitative estimate of its size within the BO framework is nevertheless presented in Sec.~\ref{sec:other}, and we take this restriction into account when interpreting the near-threshold results of Sec.~\ref{sec:thresholds}.

The dynamical diquark model, originally proposed in Ref.~\cite{Brodsky:2014xia}, postulates that the confinement mechanism itself is responsible for the binding of exotic states. In a hard production process, a diquark-antidiquark pair is formed essentially instantaneously and, as a consequence of the kinematics of the production process, rapidly separates. Being colored objects, the diquark $\delta$ and antidiquark $\bar{\delta}$ cannot move arbitrarily far apart; instead, a color flux tube (or string) develops between them, and the system evolves in a way that is closely analogous to heavy quarkonium. The model was subsequently refined and extended in Refs.~\cite{Lebed:2017min,Giron:2019bcs,Giron:2019cfc,Giron:2020qpb,Giron:2020wpx}, where $\delta$ and $\bar{\delta}$ are treated as quasi-bound hadronic subcomponents whose relative motion is quantized in the BO potentials introduced above. The distinct quantum numbers of the gluonic field give rise to towers of states labelled by the BO quantum numbers $\Lambda_\eta^\epsilon$ familiar from diatomic molecular physics, with the lowest multiplet supported by the ground-state potential $\Sigma_g^+$.

The dynamical diquark model assumes that the diquarks are point-like and carry no internal orbital excitation, so that the entire quantum-number content of a given multiplet is encoded in the diquark and antidiquark spins together with the relative orbital angular momentum $L$ and gluonic excitation. For the lowest $S$-wave multiplet $\Sigma_g^+(1S)$, in the basis of good diquark-spin eigenvalues, the states of the $\delta\text{-}\bar{\delta}$ system are~\cite{Lebed:2017min,Giron:2019cfc}
\begin{align}
J^{PC} = 0^{++}:\quad
& X_0 \equiv |0_\delta,0_{\bar{\delta}}\rangle_0,\quad
  X_0' \equiv |1_\delta,1_{\bar{\delta}}\rangle_0, \nonumber\\
J^{PC} = 1^{++}:\quad
& X_1 \equiv \tfrac{1}{\sqrt{2}}\!\left(|1_\delta,0_{\bar{\delta}}\rangle_1 + |0_\delta,1_{\bar{\delta}}\rangle_1\right), \nonumber\\
J^{PC} = 1^{+-}:\quad
& Z \equiv \tfrac{1}{\sqrt{2}}\!\left(|1_\delta,0_{\bar{\delta}}\rangle_1 - |0_\delta,1_{\bar{\delta}}\rangle_1\right), \nonumber\\
& Z' \equiv |1_\delta,1_{\bar{\delta}}\rangle_1, \nonumber\\
J^{PC} = 2^{++}:\quad
& X_2 \equiv |1_\delta,1_{\bar{\delta}}\rangle_2,
\label{eq:diquarkstates}
\end{align}
where the inner subscripts denote the spins of $\delta$ and $\bar{\delta}$ and the outer subscript gives the total spin $J$, which for $S$-wave states coincides with the total quark spin $S$. These six basis states span the full $\Sigma_g^+(1S)$ multiplet. When the two quarks inside $\delta$ (or the two antiquarks inside $\bar{\delta}$) are identical, Fermi statistics, together with the antisymmetry of the color-$\bar{\mathbf{3}}_c$ configuration and the symmetry of the $S$-wave spatial wave function, forces the spin wave function to be symmetric, leaving only the $s_\delta=1$ (or $s_{\bar{\delta}}=1$) configuration~\cite{Giron:2020wpx}. This restriction applies, for example, to $cc$ and $\bar{b}\bar{b}$ clusters, but not to heterogeneous clusters such as $[ud]$ or $[\bar{b}\bar{c}]$, where both spin-$0$ and spin-$1$ configurations are allowed.

The $T_{bc}$ system of interest is an \emph{open-flavor} tetraquark, built from a light diquark $\delta=[ud]$ and an antidiquark $\bar{\delta}'=[\bar{b}\bar{c}]$ which are \emph{not} charge-conjugate partners. Several consequences follow immediately. First, since the two heavy antiquarks carry different flavors, the Pauli principle does not constrain the $\bar{b}\bar{c}$ cluster, and both the scalar ($s_{\bar{\delta}'}=0$) and axial-vector ($s_{\bar{\delta}'}=1$) configurations are allowed. This is in marked contrast to the doubly-bottom $ud\bar{b}\bar{b}$ system, in which only the $s_{\bar{\delta}}=1$ configuration survives in the $S$-wave. Second, because $\delta$ and $\bar{\delta}'$ are no longer $C$-conjugate partners, $C$ parity is not a good quantum number for individual $T_{bc}$ states, and the symmetric and antisymmetric combinations appearing in the $X_1$ and $Z$ states of Eq.~\eqref{eq:diquarkstates} do not correspond to distinct physical eigenstates~\cite{Mutuk:2025hql}. The labels $J^{PC}$ in Eq.~\eqref{eq:diquarkstates} should therefore be understood only as a convenient shorthand for the underlying spin couplings.

The appropriate basis for the open-flavor system is the direct-product basis $|s_\delta,s_{\bar{\delta}'}\rangle_J$, with $s_\delta,s_{\bar{\delta}'}\in\{0,1\}$ and $J$ the total spin of the tetraquark. The full $\Sigma_g^+(1S)$ multiplet is then spanned by the six states
\begin{align}
&|0_\delta,0_{\bar{\delta}'}\rangle_{J=0}, \quad |1_\delta,1_{\bar{\delta}'}\rangle_{J=0}, \nonumber\\
&|1_\delta,0_{\bar{\delta}'}\rangle_{J=1}, \quad |0_\delta,1_{\bar{\delta}'}\rangle_{J=1}, \quad |1_\delta,1_{\bar{\delta}'}\rangle_{J=1}, \nonumber\\
&|1_\delta,1_{\bar{\delta}'}\rangle_{J=2},
\label{eq:bcud_states}
\end{align}
yielding two $J^P=0^+$, three $J^P=1^+$, and one $J^P=2^+$ states, all with $P=(-1)^L=+1$.

A crucial structural feature of the $T_{bc}$ system is that Fermi statistics for the $S$-wave, color-$\bar{\mathbf{3}}_c$ light diquark locks the isospin of the state to the spin of the light diquark. Because the combined color $\otimes$ space $\otimes$ flavor $\otimes$ spin wave function must be totally antisymmetric, and color is antisymmetric while $L=0$ space is symmetric, the product of flavor and spin must be antisymmetric. This gives the rigid correlation
\begin{equation}
I=0 \;\Leftrightarrow\; s_\delta=0 \ (\text{good}),
\qquad
I=1 \;\Leftrightarrow\; s_\delta=1 \ (\text{bad}),
\label{eq:IsospinSpinLock}
\end{equation}
with the combinations $(I,s_\delta)=(0,1)$ and $(1,0)$ forbidden. As a consequence, the physical Hilbert space of the light diquark is two-dimensional rather than four-dimensional: $I$ and $s_\delta$ are \emph{not} independent quantum numbers but are rigidly tied together.

Imposing Eq.~\eqref{eq:IsospinSpinLock} on the six states of Eq.~\eqref{eq:bcud_states}, the $\Sigma_g^+(1S)$ multiplet splits into two physically distinct sectors. The isoscalar sector is built on the good light diquark ($s_\delta=0$), which can combine with either a scalar or an axial-vector heavy antidiquark to give
\begin{align}
T_{bc}^{(0)} &\equiv \bigl|[ud],\,[\bar{b}\bar{c}]\bigr\rangle_{J=0},
   & I(J^P) &= 0(0^+), \label{eq:T0}\\[2pt]
T_{bc}^{(1)} &\equiv \bigl|[ud],\,\{\bar{b}\bar{c}\}\bigr\rangle_{J=1},
   & I(J^P) &= 0(1^+). \label{eq:T1}
\end{align}
These are the only two $I=0$ members of the multiplet; in particular, the state $|0_\delta,1_{\bar{\delta}'}\rangle_{J=1}$ appearing in Eq.~\eqref{eq:bcud_states} is identical to $T_{bc}^{(1)}$, while $|1_\delta,0_{\bar{\delta}'}\rangle_{J=1}$ has $s_\delta=1$ and therefore belongs to the isovector sector rather than to $I=0$. The isovector sector, built on the bad light diquark ($s_\delta=1$), then contains only configurations with $s_{\bar{\delta}'}=1$, namely
\begin{align}
\widetilde{T}_{bc}^{(0)} &\equiv \bigl|\{ud\},\,\{\bar{b}\bar{c}\}\bigr\rangle_{J=0},
   & I(J^P) &= 1(0^+), \label{eq:T0tilde}\\[2pt]
\widetilde{T}_{bc}^{(1)} &\equiv \bigl|\{ud\},\,\{\bar{b}\bar{c}\}\bigr\rangle_{J=1},
   & I(J^P) &= 1(1^+), \label{eq:T1tilde}\\[2pt]
\widetilde{T}_{bc}^{(2)} &\equiv \bigl|\{ud\},\,\{\bar{b}\bar{c}\}\bigr\rangle_{J=2},
   & I(J^P) &= 1(2^+), \label{eq:T2tilde}
\end{align}
together with a fourth isovector state,
\begin{equation}
\widetilde{T}_{bc}^{\prime(1)} \equiv \bigl|\{ud\},\,[\bar{b}\bar{c}]\bigr\rangle_{J=1},
\qquad I(J^P)=1(1^+),
\label{eq:T1tildeprime}
\end{equation}
which is built on the \emph{scalar} heavy antidiquark. The two isovector $1^{+}$ states are therefore not degenerate: as shown in Sec.~\ref{sec:hamiltonian}, they are separated by exactly the same combination of the heavy-antidiquark mass gap and chromomagnetic coupling that splits the two isoscalar states. Altogether, the $I=0$ sector contains two states, and the $I=1$ sector contains four, for a total of six, in agreement with the dimensionality of the $\Sigma_g^+(1S)$ multiplet.

The scalar state $T_{bc}^{(0)}$ is the natural dynamical-diquark candidate for the near-threshold $I(J^P)=0(0^+)$ pole found in the $B$-$\bar{D}$ scattering analysis of Ref.~\cite{Alexandrou:2023cqg}, while the axial-vector state $T_{bc}^{(1)}$ is the counterpart of the $T_{bc}$ bound state reported on the lattice in Refs.~\cite{Francis:2018jyb,Padmanath:2023rdu,Alexandrou:2023cqg}. The isoscalar states are the primary focus of the present work, since they are the ones for which lattice-QCD evidence exists and which lie in the immediate vicinity of the lowest open-flavor thresholds. The isovector members of the multiplet are nevertheless computed and discussed in Sec.~\ref{sec:other}, where they are found to lie substantially higher in mass---above every $B^{(*)}\bar{D}^{(*)}$ threshold---and are therefore expected to be broad, consistent with the absence of any lattice-QCD signal for them so far.

\subsection{Effective Hamiltonian}
\label{sec:hamiltonian}

Following the dynamical-diquark analyses of open-flavor systems in Refs.~\cite{Giron:2019cfc,Giron:2020qpb,Mutuk:2025hql}, and specializing to the $S$-wave $\Sigma_g^+(1S)$ multiplet relevant for the $T_{bc}$ ground states, the effective Hamiltonian of the $\delta$--$\bar{\delta}'$ system is written as
\begin{equation}
H = H_0 + H_\kappa,
\label{eq:H_eff}
\end{equation}
with
\begin{equation}
H_0 = \frac{\mathbf{p}^2}{2\mu} + V_{\Sigma_g^+}(r),
\qquad
\mu = \frac{m_\delta\,m_{\bar{\delta}'}}{m_\delta + m_{\bar{\delta}'}}.
\label{eq:H0}
\end{equation}
Here $V_{\Sigma_g^+}(r)$ is the BO potential between the color-$\bar{\mathbf{3}}_c$ diquark $\delta=[ud]$ and the color-$\mathbf{3}_c$ antidiquark $\bar{\delta}'=[\bar{b}\bar{c}]$, taken from lattice-QCD determinations of the static potential in the fundamental representation~\cite{Juge:2002br,Juge:1999ie}. Since we restrict ourselves to $L=0$, all spin-orbit and tensor contributions vanish identically, and the remaining spin-dependent dynamics is contained in $H_\kappa$.

Because $T_{bc}$ is an open-flavor system with different content in the two clusters, the spin-spin Hamiltonian contains two independent short-range color-magnetic couplings:
\begin{equation}
H_\kappa \;=\; 2\kappa_{ud}\,(\mathbf{s}_u \cdot \mathbf{s}_d) \;+\; 2\kappa_{\bar{b}\bar{c}}\,(\mathbf{s}_{\bar{b}} \cdot \mathbf{s}_{\bar{c}}),
\label{eq:Hkappa}
\end{equation}
where $\kappa_{ud}$ and $\kappa_{\bar{b}\bar{c}}$ describe the color-magnetic interactions within the light and heavy clusters, respectively. Their expectation values in the diquark-spin basis $|s_\delta,s_{\bar{\delta}'}\rangle_J$ follow from
\begin{equation}
\langle \mathbf{s}_q\cdot\mathbf{s}_{q'}\rangle \;=\;
\tfrac{1}{2}\!\left[s(s+1) - \tfrac{3}{2}\right] \;=\;
\begin{cases}
-3/4, & s=0,\\
+1/4, & s=1,
\end{cases}
\label{eq:ss_expval}
\end{equation}
applied separately to each cluster. Combined with the isospin--spin lock of Eq.~\eqref{eq:IsospinSpinLock}, these matrix elements automatically generate the full isoscalar--isovector splitting of the $\Sigma_g^+(1S)$ multiplet without the need for any additional isospin-dependent operator.

The physical mass of a $T_{bc}$ tetraquark state $|n;s_\delta,s_{\bar{\delta}'};J\rangle$ with radial quantum number $n$, diquark spins $(s_\delta,s_{\bar{\delta}'})$, and total spin $J$ is given by
\begin{equation}
M \;=\; m_\delta + m_{\bar{\delta}'} + E_n + \langle H_\kappa\rangle,
\label{eq:mass_full}
\end{equation}
where $E_n$ is the eigenvalue of the radial Schr\"odinger equation,
\begin{equation}
\!\left[-\frac{1}{2\mu r^2}\frac{d}{dr}\!\left(r^2\frac{d}{dr}\right) + V_{\Sigma_g^+}(r)\right]\!\psi_n(r) = E_n\,\psi_n(r),
\label{eq:schroedinger}
\end{equation}
with $L=0$ imposed throughout. Eq.~\eqref{eq:schroedinger} is solved numerically on a discretized radial grid with Dirichlet boundary conditions $\psi(0)=\psi(\infty)=0$, using a shooting algorithm that matches the logarithmic derivative of the wave function with an adaptive step size to ensure numerical stability in the strongly confining regime.

A remark on terminology is in order before proceeding. The quantity $M$ in Eq.~\eqref{eq:mass_full} is the eigenvalue of the single-channel Hamiltonian of Eq.~\eqref{eq:H_eff}, in which the confining $\Sigma_g^+$ potential supports a purely discrete spectrum; it is \emph{not} the position of a pole of the physical $B^{(*)}\bar{D}^{(*)}$ scattering amplitude. Rigorously distinguishing a bound state, a virtual state, and a resonance requires the analytic continuation of a scattering amplitude across the relevant thresholds, which the present framework does not provide. Accordingly, when in the following we compare $M$ with two-meson thresholds and refer to a state as a ``bound-state candidate'' or a ``resonance candidate,'' these designations indicate only where the single-channel eigenvalue falls relative to the threshold in question, together with the interpretation that this location suggests; whether the corresponding pole of the full amplitude is genuinely bound, virtual, or resonant---and on which Riemann sheet it resides---can only be decided by a coupled-channel scattering analysis that is beyond the scope of this work.

We characterize the spatial structure of the $\delta$--$\bar{\delta}'$ wave functions through two complementary observables: the mean separation $\langle r\rangle$ and the inverse-mean-radius $\langle 1/r\rangle^{-1}$. These two quantities probe different length scales of the bound state and, together, provide a more discriminating picture than a single radius can. The expectation value $\langle r\rangle$ is sensitive to the long-distance behavior of the wave function and thus reflects the extent of the confining color flux tube that connects the diquark and antidiquark. In contrast, $\langle 1/r\rangle^{-1}$ is weighted toward small separations and therefore probes the short-distance region where the interaction is dominated by attractive physics. This distinction is particularly informative for a Cornell-type potential of the form $V(r) = -\alpha/r + \sigma r$, in which the color-Coulomb term $-\alpha/r$ governs the short-range dynamics and the linear term $\sigma r$ is responsible for long-range confinement. In such a setup, $\langle 1/r\rangle^{-1}$ provides a direct measure of the typical strength of the short-range attraction felt by the cluster pair, while $\langle r\rangle$ characterizes the overall size of the confined system. The combination of the two is especially useful for distinguishing a compact diquark-antidiquark configuration from a loosely bound hadronic molecule: molecular states generically exhibit both a large mean separation and a large $\langle 1/r\rangle^{-1}$, with both quantities typically of order $1$~fm or more, whereas compact $\delta\bar{\delta}'$ states have $\langle 1/r\rangle^{-1}$ well below the hadronic scale, signaling genuine short-range overlap between the clusters.

Combining Eqs.~\eqref{eq:Hkappa}, \eqref{eq:ss_expval}, and \eqref{eq:mass_full} with the state definitions in Eqs.~\eqref{eq:T0}--\eqref{eq:T2tilde}, the physical masses of the isoscalar ground states $T_{bc}^{(0)}$ and $T_{bc}^{(1)}$ take the closed-form expressions
\begin{align}
M\!\left(T_{bc}^{(0)}\right) &= m_{[ud]} + m_{[\bar{b}\bar{c}]} + E_{0}^{(0)} - \tfrac{3}{2}\kappa_{ud} - \tfrac{3}{2}\kappa_{\bar{b}\bar{c}}, \label{eq:M_T0}\\[2pt]
M\!\left(T_{bc}^{(1)}\right) &= m_{[ud]} + m_{\{\bar{b}\bar{c}\}} + E_{0}^{(1)} - \tfrac{3}{2}\kappa_{ud} + \tfrac{1}{2}\kappa_{\bar{b}\bar{c}}, \label{eq:M_T1}
\end{align}
where $m_{[\bar{b}\bar{c}]}$ and $m_{\{\bar{b}\bar{c}\}}$ are, respectively, the antisymmetric (scalar) and symmetric (axial-vector) heavy-antidiquark masses, and $E_{0}^{(0,1)}$ denotes the BO ground-state eigenvalue computed with the corresponding reduced mass $\mu^{(0,1)}$. The corresponding isovector masses are
\begin{align}
M\!\left(\widetilde{T}_{bc}^{(0)}\right) \;=&\; M\!\left(\widetilde{T}_{bc}^{(1)}\right) \;=\; M\!\left(\widetilde{T}_{bc}^{(2)}\right) \nonumber\\
=&\; m_{[ud]} + m_{\{\bar{b}\bar{c}\}} + E_{0}^{(1)} \nonumber\\
&+ \tfrac{1}{2}\kappa_{ud} + \tfrac{1}{2}\kappa_{\bar{b}\bar{c}}, \label{eq:M_Ttilde}
\end{align}
while the fourth isovector state of Eq.~\eqref{eq:T1tildeprime}, which is built on the scalar heavy antidiquark, has
\begin{align}
M\!\left(\widetilde{T}_{bc}^{\prime(1)}\right) \;=&\; m_{[ud]} + m_{[\bar{b}\bar{c}]} + E_{0}^{(0)} \nonumber\\
&+ \tfrac{1}{2}\kappa_{ud} - \tfrac{3}{2}\kappa_{\bar{b}\bar{c}}. \label{eq:M_Ttildeprime}
\end{align}
The threefold degeneracy of the $\widetilde{T}_{bc}^{(J)}$ states is a direct consequence of the fact that all three are built on the same $|1_{[ud]},1_{\{\bar{b}\bar{c}\}}\rangle$ internal configuration and differ only in how the two spin-$1$ clusters are coupled to total $J=0,1,2$. Since $H_\kappa$ of Eq.~\eqref{eq:Hkappa} acts only on the internal spins of the individual clusters and is blind to the way in which the cluster spins are subsequently combined, it assigns the same eigenvalue to all three isovector members. This degeneracy is lifted only by spin-orbit, tensor, and higher-order interactions that are not included in the present analysis. The state $\widetilde{T}_{bc}^{\prime(1)}$, by contrast, is \emph{not} degenerate with them: subtracting Eq.~\eqref{eq:M_Ttildeprime} from Eq.~\eqref{eq:M_Ttilde} gives $\bigl(m_{\{\bar{b}\bar{c}\}}-m_{[\bar{b}\bar{c}]}\bigr) + \bigl(E_{0}^{(1)}-E_{0}^{(0)}\bigr) + 2\kappa_{\bar{b}\bar{c}}$, i.e.\ precisely the isoscalar hyperfine splitting of Eq.~\eqref{eq:JP_splitting}, so that the two $I=1$, $J^P=1^+$ levels repeat the $0^+$--$1^+$ pattern of the isoscalar sector. Since both carry the same conserved quantum numbers, interactions beyond $H_\kappa$ will in general mix them; such mixing is not included here.

Two important mass splittings follow immediately from Eqs.~\eqref{eq:M_T0}--\eqref{eq:M_Ttilde}. Within the isoscalar sector, the axial-vector--scalar splitting receives a kinematic contribution from the heavy-antidiquark mass gap and a chromomagnetic contribution from the heavy-antidiquark spin-spin coupling,
\begin{align}
\Delta_{HF}\;\equiv&\;M\!\left(T_{bc}^{(1)}\right) - M\!\left(T_{bc}^{(0)}\right) \nonumber\\
=&\;\bigl(m_{\{\bar{b}\bar{c}\}}-m_{[\bar{b}\bar{c}]}\bigr) \nonumber\\
&+ \bigl(E_{0}^{(1)}-E_{0}^{(0)}\bigr) + 2\,\kappa_{\bar{b}\bar{c}}.
\label{eq:JP_splitting}
\end{align}
For the inputs adopted in Sec.~\ref{sec:results}, the kinematic gap dominates ($m_{\{\bar{b}\bar{c}\}}-m_{[\bar{b}\bar{c}]}\simeq 40$~MeV), the BO contribution $E_{0}^{(1)}-E_{0}^{(0)}$ is at the sub-MeV level because the two reduced masses are very close, and the chromomagnetic term $2\kappa_{\bar{b}\bar{c}}\in[20,40]$~MeV. The slope $\partial\Delta_{HF}/\partial\kappa_{\bar{b}\bar{c}}=2$ is therefore the only quantity that responds to a variation of the chromomagnetic coupling at fixed kinematic input, and a measurement of $\Delta_{HF}$ determines $\kappa_{\bar{b}\bar{c}}$ only after the heavy-antidiquark mass gap has been independently fixed.
The isoscalar--isovector gap, on the other hand, depends on which $J^P$ channel is compared. Taking the two $J^P=1^+$ states, one finds
\begin{equation}
M\!\left(\widetilde{T}_{bc}^{(1)}\right) - M\!\left(T_{bc}^{(1)}\right) \;=\; 2\,\kappa_{ud},
\label{eq:isogap_1p}
\end{equation}
which is controlled purely by the light-diquark spin-spin coupling and directly reflects the energy cost of promoting the light $[ud]$ pair from the good to the bad configuration; the heavy-antidiquark mass and BO contributions cancel because both states are built on the symmetric $\{\bar{b}\bar{c}\}$ cluster. Comparing instead the two $J^P=0^+$ states gives
\begin{align}
&M\!\left(\widetilde{T}_{bc}^{(0)}\right) - M\!\left(T_{bc}^{(0)}\right) \nonumber\\
&\quad=\;\bigl(m_{\{\bar{b}\bar{c}\}}-m_{[\bar{b}\bar{c}]}\bigr) + \bigl(E_{0}^{(1)}-E_{0}^{(0)}\bigr) \nonumber\\
&\qquad + 2\,\kappa_{ud} + 2\,\kappa_{\bar{b}\bar{c}},
\label{eq:isogap_0p}
\end{align}
because the isoscalar $0^+$ is built on a scalar heavy antidiquark while the isovector $0^+$ is built on an axial-vector one, so that the gap accumulates the antisymmetric--symmetric antidiquark mass difference together with the spin-spin cost of both cluster transitions. For the parameter ranges adopted in Sec.~\ref{sec:results}, Eqs.~\eqref{eq:isogap_1p} and \eqref{eq:isogap_0p} yield isoscalar--isovector gaps of $\simeq 2\kappa_{ud}\approx 206$~MeV and $\simeq 40+2\kappa_{ud}+2\kappa_{\bar{b}\bar{c}}\approx 266$--$286$~MeV, respectively, indicating that the isoscalar channels are by far the most favorable for the formation of strong-interaction-stable $T_{bc}$ tetraquarks. The diquark masses $m_{[ud]}$ and $m_{[\bar{b}\bar{c}]}$, as well as the spin-spin couplings $\kappa_{ud}$ and $\kappa_{\bar{b}\bar{c}}$, are taken as external inputs from QCD sum rules and meson phenomenology, as detailed in Sec.~\ref{sec:results}.

\section{Numerical Results}
\label{sec:results}

\subsection{Model inputs and the consistency of their sources}
\label{sec:inputs}

We now present the numerical results of the dynamical diquark model for the $T_{bc}$ system in the $\Sigma_g^+(1S)$ multiplet.
For the light diquark we adopt $m_{[ud]} = 0.64 \pm 0.06$~GeV from the QCD sum-rule analysis \cite{Wang:2011ab}. For the heavy
antidiquark we use $m_{[\bar{b}\bar{c}]} = 6.38$~GeV and $m_{\{\bar{b}\bar{c}\}} = 6.42$~GeV, obtained from the Regge-trajectory
approach \cite{Feng:2023txx}. For the light-diquark effective spin--spin coupling we take $\kappa_{ud} = 103$~MeV, as determined from the $\Sigma_Q$--$\Lambda_Q$ baryon splittings in the diquark--antidiquark analysis in Ref.~\cite{Maiani:2004vq}. To probe the sensitivity of the spectrum to the heavy-antidiquark spin--spin interaction, we consider three representative parameter sets, labeled Set~I, Set~II, and
Set~III, corresponding respectively to $\kappa_{\bar{b}\bar{c}} = 10$, $15$, and $20$~MeV:
\begin{align}
\text{Set I:}\quad   & \kappa_{\bar{b}\bar{c}} = 10~\text{MeV},\\[2pt]
\text{Set II:}\quad  & \kappa_{\bar{b}\bar{c}} = 15~\text{MeV},\\[2pt]
\text{Set III:}\quad & \kappa_{\bar{b}\bar{c}} = 20~\text{MeV}.
\end{align}
All three values lie well within the phenomenological range mentioned by the QCD Laplace sum-rules analysis \cite{Esau:2019hqw}, and the
light-diquark mass, the heavy-antidiquark masses, and the light-diquark spin--spin coupling $\kappa_{ud}$ are held fixed across all three sets.
The BO potential $V_{\Sigma_g^+}(r)$ is taken from the lattice-QCD parametrizations of Refs.~\cite{Juge:2002br,Juge:1999ie}, and the radial Schr\"odinger equation is solved numerically as described in Sec.~\ref{sec:method}. For ease of reference, Table~\ref{tab:inputs} collects all model inputs together with their adopted values, sources, quoted uncertainties, and the role they play in the calculation.

\begin{table*}[t]
\centering
\caption{Model inputs used in this work: adopted value, source, quoted uncertainty, and role in the calculation. The inputs are taken from independent determinations (QCD sum rules, Regge phenomenology, baryon spectroscopy, and lattice QCD); they are effective parameters defined within their respective frameworks, and residual scheme dependence among them is included in the systematic uncertainty discussed in the text.}
\label{tab:inputs}
\renewcommand{\arraystretch}{1.2}
\small
\begin{tabular}{lcccc} \hline
Input & Value & Source & Uncertainty & Role \\ \hline
$m_{[ud]}$ & $0.64$~GeV & QCD sum rules~\cite{Wang:2011ab} & $\pm 0.06$~GeV & light-diquark mass; reduced mass \\
$m_{[\bar b\bar c]}$ & $6.38$~GeV & Regge trajectories~\cite{Feng:2023txx} & few tens of MeV & scalar heavy-antidiquark mass \\
$m_{\{\bar b\bar c\}}$ & $6.42$~GeV & Regge trajectories~\cite{Feng:2023txx} & few tens of MeV & axial-vector heavy-antidiquark mass \\
$\kappa_{ud}$ & $103$~MeV & $\Sigma_Q$--$\Lambda_Q$ splittings~\cite{Maiani:2004vq} & $\mathcal{O}(10\%)$ & light-diquark spin--spin coupling \\
$\kappa_{\bar b\bar c}$ & $10$--$20$~MeV & Laplace sum-rule range~\cite{Esau:2019hqw} & scanned (Sets I--III) & heavy-antidiquark spin--spin coupling \\
$V_{\Sigma_g^+}(r)$ & lattice parametrization & Refs.~\cite{Juge:1999ie,Juge:2002br} & $\mathcal{O}(10)$~MeV on $E_0$ & BO potential; radial eigenvalue $E_0$ \\ \hline
\end{tabular}
\end{table*}

The inputs collected in Table~\ref{tab:inputs} originate from four different frameworks---QCD sum rules, Regge phenomenology, baryon spectroscopy, and lattice QCD---and it is appropriate to explain why this is unavoidable and in what sense the resulting set is nevertheless internally consistent. The Hamiltonian of Eq.~\eqref{eq:H_eff} requires three logically distinct kinds of input. The first is the static energy of a color-$\bar{\mathbf{3}}_c$ source in the presence of a color-$\mathbf{3}_c$ antisource: this is a gauge-invariant, well-defined quantity, and lattice QCD is the only first-principles source for it. The second is the masses of the two clusters. These are \emph{not} observables---a diquark is a colored, gauge-variant object with no asymptotic state---so they can only be defined within a model or through a correlator supplemented by a gauge fixing or a Wilson line, as in the QCD sum-rule and Regge analyses employed here. The third is the short-range chromomagnetic couplings, which are defined operationally by their action on measured hadron splittings. No single calculation supplies all three, and the same division of labor is made in every application of the dynamical diquark model, in which the BO potentials are taken from lattice QCD while the diquark masses and the couplings $\kappa$ are taken from spectroscopy~\cite{Lebed:2017min,Giron:2019bcs,Giron:2019cfc,Giron:2020qpb}. Consistency in this context therefore cannot mean a common source; what it does require is a common \emph{definition}, namely that each input be exactly the quantity appearing in Eq.~\eqref{eq:H_eff}---the mass of a compact color-triplet cluster treated as a static source, and a contact spin-spin coupling between the constituents of that cluster. The determinations of Refs.~\cite{Wang:2011ab,Feng:2023txx,Maiani:2004vq,Esau:2019hqw} all satisfy this requirement, and it is in this sense that they may be combined.

A second, more technical point governs how much of the residual scheme dependence is physical. The spectrum generated by Eqs.~\eqref{eq:mass_full}--\eqref{eq:schroedinger} depends on the cluster masses and on the BO potential only through the combination $m_\delta + m_{\bar{\delta}'} + V_{\Sigma_g^+}(r)$: the substitution $V_{\Sigma_g^+}(r)\to V_{\Sigma_g^+}(r)+\Delta$ together with $m_\delta+m_{\bar{\delta}'}\to m_\delta+m_{\bar{\delta}'}-\Delta$ leaves every eigenvalue and every wave function unchanged. Constituent cluster masses are thus meaningful only relative to the additive convention adopted for the static energy---and the lattice $\Sigma_g^+$ potential carries precisely such a scheme-dependent self-energy constant. A mismatch between the conventions of two determinations consequently appears as a \emph{common} shift of the whole spectrum rather than as a distortion of it. This is the technical reason why absolute masses carry a substantially larger systematic uncertainty than splittings, and it also implies that quantities defined as differences---the chromomagnetic couplings and the cluster mass gaps---are transferable between frameworks, whereas absolute cluster masses are not. We accordingly cross-check the former, and assign a conservative uncertainty to the latter.

The least constrained input, $\kappa_{\bar b\bar c}$, can indeed be checked against a determination entirely independent of the sum rules of Ref.~\cite{Esau:2019hqw}. Because the color factor $\langle\boldsymbol{\lambda}_i\cdot\boldsymbol{\lambda}_j\rangle$ equals $-8/3$ for a $\bar{\mathbf{3}}_c$ diquark but $-16/3$ for a color-singlet meson, the chromomagnetic coupling inside a heavy antidiquark is, up to differences in the wave function at the origin, half of its mesonic counterpart,
\begin{equation}
\kappa_{\bar b\bar c}\;\simeq\;\tfrac{1}{2}\,\kappa_{b\bar c}^{\rm (meson)}
\;=\;\tfrac{1}{4}\left[M(B_c^{*})-M(B_c)\right].
\label{eq:kappa_Bc}
\end{equation}
Lattice QCD predicts $M(B_c^{*})-M(B_c)=54(3)$~MeV~\cite{Dowdall:2012ab}, while quark-model estimates extend to about $70$~MeV, giving $\kappa_{\bar b\bar c}\simeq 13$--$18$~MeV. The interval $10$--$20$~MeV scanned in Sets~I--III thus brackets this independent estimate, with Set~II close to its center; the spread of the spectrum reported below is therefore representative of the true uncertainty in this coupling rather than an arbitrary choice.

One source-related ambiguity does, however, affect a physical quantity directly, and we quantify it here. It concerns the apportionment of the heavy-antidiquark spin dependence, which in the parameter set of Table~\ref{tab:inputs} enters in two places: through the two distinct Regge masses $m_{[\bar b\bar c]}$ and $m_{\{\bar b\bar c\}}$, whose $40$~MeV gap is itself of chromomagnetic origin, and through the term $2\kappa_{\bar b\bar c}(\mathbf{s}_{\bar b}\cdot\mathbf{s}_{\bar c})$ of Eq.~\eqref{eq:Hkappa}. Two limiting conventions bracket the possibilities. In convention~(A), used for our central results, the two contributions are treated as physically distinct: the Regge gap measures the spin splitting of the \emph{isolated} $\bar b\bar c$ cluster, whereas $\kappa_{\bar b\bar c}$ describes the residual chromomagnetic interaction as it acts inside the tetraquark, where the cluster is embedded in the color field of the light diquark and its wave function at the origin is modified. In convention~(B), which is the one normally adopted in dynamical-diquark analyses~\cite{Lebed:2017min,Giron:2019cfc}, the cluster is assigned a single spin-averaged mass,
\begin{equation}
\bar{m}_{\bar b\bar c} \;=\; \tfrac{1}{4}\,m_{[\bar b\bar c]} + \tfrac{3}{4}\,m_{\{\bar b\bar c\}} \;=\; 6.41~\text{GeV},
\label{eq:mbar}
\end{equation}
and the entire spin dependence is generated by $H_\kappa$; the Regge gap is then reproduced for $2\kappa_{\bar b\bar c}=40$~MeV, i.e.\ $\kappa_{\bar b\bar c}=20$~MeV, which is exactly the value adopted in Set~III. Convention~(A) maximizes and convention~(B) minimizes the hyperfine splitting, so that the physical answer is bracketed by the two. Table~\ref{tab:conventions} compares the resulting spectra, obtained from Eqs.~\eqref{eq:M_T0}--\eqref{eq:M_T1} with the same radial eigenvalue. The absolute masses differ by at most $+30$~MeV in the $0^{+}$ channel and $-10$~MeV in the $1^{+}$ channel, well inside the parametric uncertainty established below, while $\Delta_{HF}$ ranges from $2\kappa_{\bar b\bar c}=20$--$40$~MeV in convention~(B) to $59$--$79$~MeV in convention~(A). We quote the latter as our central values, but regard $\Delta_{HF}\simeq 20$--$80$~MeV as the interval allowed by the present inputs. The qualitative picture of Sec.~\ref{sec:thresholds} is affected only mildly: in both conventions the axial-vector state lies above $B^{*}\bar{D}$ and well below $B\bar{D}^{*}$, while the scalar moves from essentially at the $B\bar{D}$ threshold in convention~(A) to $24$--$39$~MeV above it in convention~(B). Since this displacement is smaller than the uncertainty on the absolute masses, both conventions are compatible with a scalar sitting at or slightly above the $B\bar{D}$ threshold, and neither of them establishes binding.

\begin{table*}[t]
\centering
\caption{Dependence of the isoscalar spectrum on the convention used to apportion the heavy-antidiquark spin dependence between the input cluster masses and $H_\kappa$. Convention~(A) uses the two Regge masses $m_{[\bar b\bar c]}=6.38$~GeV and $m_{\{\bar b\bar c\}}=6.42$~GeV together with $\kappa_{\bar b\bar c}$ (central results of this work); convention~(B) uses the spin-averaged mass of Eq.~\eqref{eq:mbar} and generates the entire splitting from $\kappa_{\bar b\bar c}$. Masses in GeV, $\Delta_{HF}$ in MeV.}
\label{tab:conventions}
\renewcommand{\arraystretch}{1.15}
\small
\begin{tabular}{lccccccc} \hline
 & \multicolumn{3}{c}{Convention (A)} & & \multicolumn{3}{c}{Convention (B)} \\
\cline{2-4}\cline{6-8}
Set & $M(T_{bc}^{(0)})$ & $M(T_{bc}^{(1)})$ & $\Delta_{HF}$ & & $M(T_{bc}^{(0)})$ & $M(T_{bc}^{(1)})$ & $\Delta_{HF}$ \\ \hline
I   & 7.158 & 7.217 & 59 & & 7.188 & 7.208 & 20 \\
II  & 7.151 & 7.220 & 69 & & 7.180 & 7.210 & 30 \\
III & 7.143 & 7.222 & 79 & & 7.173 & 7.213 & 40 \\ \hline
\end{tabular}
\end{table*}

Before presenting the spectrum, we assess the total parametric uncertainty implied by Table~\ref{tab:inputs}, since the treatment of uncertainties determines how the near-threshold comparisons below should be read. The dominant source is the light-diquark mass: propagating $\delta m_{[ud]}=\pm 60$~MeV shifts the absolute masses of both $T_{bc}^{(0)}$ and $T_{bc}^{(1)}$ by up to $\pm 60$~MeV, the direct constituent shift being partially compensated by the induced change of the BO eigenvalue through the reduced mass. The heavy-antidiquark masses carry uncertainties at the level of a few tens of MeV typical of Regge determinations, the lattice parametrization of the BO potential contributes an estimated $\mathcal{O}(10)$~MeV through the offset and string tension entering $E_0$, and the neglected color mixing and coupled-channel effects discussed in Secs.~\ref{sec:ddm} and \ref{sec:thresholds} act at a comparable scale. A realistic total uncertainty on the \emph{absolute} masses is therefore of order $\pm 60$--$70$~MeV, far larger than the spread induced by $\kappa_{\bar b\bar c}$ alone. Crucially, however, these dominant uncertainties are strongly correlated between the two isoscalar channels: $m_{[ud]}$, $\kappa_{ud}$, and the BO potential enter $M(T_{bc}^{(0)})$ and $M(T_{bc}^{(1)})$ identically, and therefore cancel in the hyperfine splitting $\Delta_{HF}$, whose uncertainty is instead controlled by the heavy-antidiquark mass gap and by $\kappa_{\bar b\bar c}$. The robust outputs of the calculation are thus the mass \emph{differences}, the ordering of the states relative to each other and to the pattern of thresholds, and the parametric dependences; the absolute proximity of any individual eigenvalue to a given threshold at the few-MeV level is a central-value statement that should not be over-interpreted.

A related point, which deserves to be stated explicitly, concerns the comparison with physical thresholds. The framework adopted here describes only the relative motion of the $\delta$--$\bar\delta'$ pair in the BO potential; the $B^{(*)}$ and $\bar D^{(*)}$ mesons are quark--antiquark color singlets that lie outside this Hilbert space, so the model neither computes nor is calibrated to the physical meson masses. The thresholds quoted in Sec.~\ref{sec:thresholds} are fixed by the experimental PDG masses, while the tetraquark masses are built from independently determined constituent inputs. Few-MeV statements about the position of an eigenvalue relative to a threshold therefore compare quantities of different origin and are meaningful only at the level of the central values, within the parametric uncertainty budget established above; this limitation is inherited by all constituent-model analyses of near-threshold states and is made explicit here so that the reader can weigh the conclusions accordingly.

The mass spectra and characteristic spatial scales of the doubly-heavy open-flavor tetraquark states $T_{bc}^{(0)}$ ($J^{P}=0^{+}$) and $T_{bc}^{(1)}$ ($J^{P}=1^{+}$), obtained for the three parameter sets, are summarized in Tables~\ref{tab:Tbc0} and \ref{tab:Tbc1}.

\begin{table}[h]
\centering
\caption{Predicted mass $M$, mean inter-cluster separation
$\langle r \rangle$, and inverse mean radius
$\langle 1/r \rangle^{-1}$ of the scalar doubly-heavy tetraquark
$T_{bc}^{(0)}$ ($J^{P}=0^{+}$) for the three parameter sets used in
this study.}
\label{tab:Tbc0}
\renewcommand{\arraystretch}{1.15}
\begin{tabular}{cccc} \hline
Parameter Set & $M(T_{bc}^{(0)})$ [GeV] & $\langle r \rangle$ [fm]
              & $\langle 1/r \rangle^{-1}$ [fm] \\ \hline
Set I   & 7.158 & 0.45 & 0.33 \\
Set II  & 7.151 & 0.45 & 0.33 \\
Set III & 7.143 & 0.45 & 0.33 \\ \hline
\end{tabular}
\end{table}

\begin{table}[h]
\centering
\caption{Same as Table~\ref{tab:Tbc0}, but for the axial-vector
doubly-heavy tetraquark $T_{bc}^{(1)}$ ($J^{P}=1^{+}$).}
\label{tab:Tbc1}
\renewcommand{\arraystretch}{1.15}
\begin{tabular}{cccc} \hline
Parameter Set & $M(T_{bc}^{(1)})$ [GeV] & $\langle r \rangle$ [fm]
              & $\langle 1/r \rangle^{-1}$ [fm] \\ \hline
Set I   & 7.217 & 0.46 & 0.34 \\
Set II  & 7.220 & 0.46 & 0.34 \\
Set III & 7.222 & 0.46 & 0.34 \\ \hline
\end{tabular}
\end{table}

\subsection{Spectrum, hyperfine splitting, and internal structure}

Across the three parameter sets, the scalar state is predicted in the narrow window $M(T_{bc}^{(0)}) = 7.143\text{--}7.158$~GeV, while the
axial-vector state lies in $M(T_{bc}^{(1)}) = 7.217\text{--}7.222$~GeV. The total spread induced by varying $\kappa_{\bar{b}\bar{c}}$ over the
range $10$--$20$~MeV is only $\Delta M \simeq 15$~MeV in the $0^{+}$ channel and $\Delta M \simeq 5$~MeV in the $1^{+}$ channel. This mild sensitivity is expected on general grounds: in the dynamical diquark picture the tetraquark mass is dominated by the constituent masses $m_{[\bar{b}\bar{c}]}$, $m_{\{\bar{b}\bar{c}\}}$, and $m_{[ud]}$ and by the BO potential $V_{\Sigma_g^+}(r)$, all of which are kept fixed; $\kappa_{\bar{b}\bar{c}}$ enters only through the heavy-antidiquark chromomagnetic term $\kappa_{\bar{b}\bar{c}}\, \vec{S}_{\bar b}\!\cdot\!\vec{S}_{\bar c}$, which is diagonal in the heavy-antidiquark spin basis.

The directions of the two trends---$M(T_{bc}^{(0)})$ decreasing and
$M(T_{bc}^{(1)})$ increasing as $\kappa_{\bar{b}\bar{c}}$ grows---are
therefore physically transparent. The scalar $T_{bc}^{(0)}$ is built
on the spin-$0$ antisymmetric antidiquark $[\bar{b}\bar{c}]$, for which
$\langle \vec{S}_{\bar b}\!\cdot\!\vec{S}_{\bar c}\rangle = -3/4$,
whereas the axial-vector $T_{bc}^{(1)}$ contains the spin-$1$
symmetric antidiquark $\{\bar{b}\bar{c}\}$, for which
$\langle \vec{S}_{\bar b}\!\cdot\!\vec{S}_{\bar c}\rangle = +1/4$. An
increase in $\kappa_{\bar{b}\bar{c}}$ thus lowers the $0^{+}$ mass by
$3\kappa_{\bar{b}\bar{c}}/2$ and raises the $1^{+}$ mass by
$\kappa_{\bar{b}\bar{c}}/2$ relative to the spin-independent baseline,
in quantitative agreement with the results of
Tables~\ref{tab:Tbc0}--\ref{tab:Tbc1}. The asymmetry between the two
shifts---a factor of three stronger in the scalar channel---directly
reflects the different Casimirs of the two antidiquark configurations
and explains why Set~III exhibits the largest hyperfine splitting
while keeping $\langle r\rangle$ essentially unchanged. Defining
$\Delta_{HF}\equiv M(T_{bc}^{(1)})-M(T_{bc}^{(0)})$, we obtain
$\Delta_{HF} \simeq 59$, $69$, and $79$~MeV for Sets I, II, and III,
respectively. The dominant contribution to the absolute value of $\Delta_{HF}$
is the $\sim 40$~MeV mass gap between the symmetric and antisymmetric
heavy-antidiquark configurations, $m_{\{\bar{b}\bar{c}\}}-m_{[\bar{b}\bar{c}]}$,
which is independent of $\kappa_{\bar{b}\bar{c}}$; the chromomagnetic coupling
adds the term $2\kappa_{\bar{b}\bar{c}}$ on top, so that each $5$~MeV increment
in $\kappa_{\bar{b}\bar{c}}$ generates the expected $10$~MeV shift in
$\Delta_{HF}$ to within a fraction of an MeV, in agreement with the analytic
slope $\partial\Delta_{HF}/\partial\kappa_{\bar{b}\bar{c}}=2$ derived from
Eq.~\eqref{eq:JP_splitting}. Once the heavy-antidiquark mass gap has been
fixed by independent input, this linear response to $\kappa_{\bar{b}\bar{c}}$
provides a clean handle with which future experimental data can be used to pin
down the chromomagnetic coupling in a regime that has so far been
inaccessible.

The two radial observables reported in the tables encode complementary
information on the internal structure. The mean inter-cluster
separation $\langle r\rangle$ characterizes the average size of the
heavy-antidiquark--light-diquark configuration, weighted by the
long-distance tail of the radial probability density
$r^{2}|\psi(r)|^{2}$. The inverse mean radius
$\langle 1/r\rangle^{-1}$ is most sensitive to the short-distance
bulk of the wave function, where the Coulomb-like one-gluon exchange
dominates. Both quantities are independent of the parameter set at
the displayed precision, reflecting the fact that
$\kappa_{\bar{b}\bar{c}}\langle \vec{S}\cdot\vec{S}\rangle\sim
\mathcal{O}(20~\text{MeV})$ is a genuinely small perturbation
compared with the typical kinetic and potential scales of the
$\Sigma_g^+(1S)$ state, and that, in the BO
formulation adopted here, the chromomagnetic interaction enters the
radial Schr\"odinger equation only as an additive constant, leaving
the wave function itself unaffected at this level.

At the precision at which they are displayed, the two channels differ by one unit in the last quoted digit,
\begin{equation}
\langle r\rangle^{(0)} = 0.45~\text{fm},\quad
\langle r\rangle^{(1)} = 0.46~\text{fm},
\end{equation}
\begin{equation}
\langle 1/r\rangle^{-1\,(0)} = 0.33~\text{fm},\quad
\langle 1/r\rangle^{-1\,(1)} = 0.34~\text{fm},
\end{equation}
but this difference should not be assigned physical significance, as
we now show. The only quantity that distinguishes the two radial
problems is the reduced mass entering the radial Schr\"odinger
equation,
\begin{align}
\mu^{(0)}_{\bar{b}\bar{c},ud}
&= \frac{m_{[\bar{b}\bar{c}]}\, m_{[ud]}}
{m_{[\bar{b}\bar{c}]} + m_{[ud]}} \approx 0.5817~\text{GeV}, \\
\mu^{(1)}_{\bar{b}\bar{c},ud}
&= \frac{m_{\{\bar{b}\bar{c}\}}\, m_{[ud]}}
{m_{\{\bar{b}\bar{c}\}} + m_{[ud]}} \approx 0.5820~\text{GeV},
\end{align}
where the antisymmetric and symmetric heavy-antidiquark masses,
$m_{[\bar{b}\bar{c}]}=6.38$~GeV and $m_{\{\bar{b}\bar{c}\}}=6.42$~GeV
respectively, differ by $40$~MeV. Because the light diquark is more
than an order of magnitude lighter than the heavy antidiquark, this
$40$~MeV gap translates into a fractional change of the reduced mass
of only $\Delta\mu/\mu \simeq 5\times 10^{-4}$. Moreover, both pieces
of the BO potential produce a size that \emph{decreases}
monotonically with increasing $\mu$---as $a_0\sim 1/(\mu\alpha)$ in
the Coulombic limit and as $\langle r\rangle \sim
(\mu\sigma)^{-1/3}$ in the linear one---so that the axial-vector
state, which carries the marginally larger reduced mass, is expected
to be very slightly \emph{more} compact than the scalar one, by a
relative amount of order $10^{-4}$, i.e.\ by some $10^{-4}$~fm. This
is three orders of magnitude below the two-digit precision at which
the radii are quoted in
Tables~\ref{tab:Tbc0}--\ref{tab:Tbc1}. The single-unit difference in
the last digit is therefore a rounding artifact of the numerical
grid, and carries neither a physical magnitude nor a physical sign.
The meaningful statement is that the two states share, to the
accuracy of the present calculation, one and the same
$\Sigma_g^+(1S)$ radial wave function: the chromomagnetic term enters
Eq.~\eqref{eq:schroedinger} only as an additive constant, and the
heavy-antidiquark mass gap is negligible on the scale of the reduced
mass.

The two states share an essentially common shape, as quantified by
the dimensionless ratio
\begin{align}
\frac{\langle 1/r\rangle^{-1}}{\langle r\rangle}
&\simeq 0.73\;\text{(scalar)}, \\
\frac{\langle 1/r\rangle^{-1}}{\langle r\rangle}
&\simeq 0.74\;\text{(axial-vector)},
\end{align}
which by Jensen's inequality\footnote{Jensen's inequality states
that for any convex function $f$ (i.e., $f''\geq 0$) and any
normalized probability distribution,
$\langle f(x)\rangle \geq f(\langle x\rangle)$.  Applying this to
the convex function $f(r)=1/r$ ($r>0$) with the radial probability
density $|\psi(r)|^{2}r^{2}$ gives
$\langle 1/r\rangle \geq 1/\langle r\rangle$, or equivalently
$\langle 1/r\rangle^{-1}\leq\langle r\rangle$.  Equality holds if
and only if $|\psi(r)|^{2}$ is a Dirac delta function, corresponding
to a state with zero spatial spread.} applied to the convex function
$1/r$ is bounded above by unity (saturated only for a delta-like
distribution).  Both values lie squarely between the pure Coulombic
($\sim 2/3$) and pure harmonic-oscillator ($\sim 0.85$) regimes,
indicating that, within the present model, the short-range gluon
exchange and the long-range confining string contribute on comparable
footings to the binding in both $J^{P}$ sectors---a pattern familiar
from heavy quarkonium phenomenology and found here to extend to the
diquark--antidiquark description of four-quark configurations.

The compactness of the wave functions---$\langle r\rangle\simeq
0.45$--$0.46$~fm, significantly smaller than the typical confinement scale
 ($\sim 1.0$~fm) and well below the deuteron-like scale
of loosely bound molecular candidates---is consistent with an
interpretation of $T_{bc}^{(0)}$ and $T_{bc}^{(1)}$ as compact
diquark--antidiquark configurations rather than extended hadronic
molecules. We emphasize, however, that this conclusion is not model
independent: the radii are computed from the wave function generated
within the assumed $\delta$--$\bar\delta'$ framework, which builds in
a compact configuration from the outset, and they cannot by
themselves provide an unambiguous discriminator between compact and
molecular structures. An extended molecular component, if present in
the physical state through the near-threshold coupling discussed
below, would simply lie outside the Hilbert space of the present
description. The radii quoted here should therefore be read as
characterizing the compact core of the state within the model, not
as a measurement of the physical size of the state.
We note in passing that within a molecular interpretation
the $J^{P}$ ordering of the radii would be governed by the
binding-energy hierarchy of the $B^{(*)}\bar D^{(*)}$ thresholds
rather than by the heavy-antidiquark mass splitting, and would be
expected to display a substantially larger relative spread than the
essentially exact degeneracy of the two radial profiles found here. The
near-coincidence of the two radial scales between the scalar and
axial-vector channels, together with their parameter-set
insensitivity, thus reinforces the picture of a common
$\Sigma_g^+(1S)$ spatial profile shared by the entire $T_{bc}$
multiplet, perturbed only marginally by the heavy-antidiquark mass
splitting and the chromomagnetic coupling.

\subsection{Thresholds and phenomenology}
\label{sec:thresholds}

\begin{figure*}[t]
\centering
\includegraphics[width=0.95\linewidth]{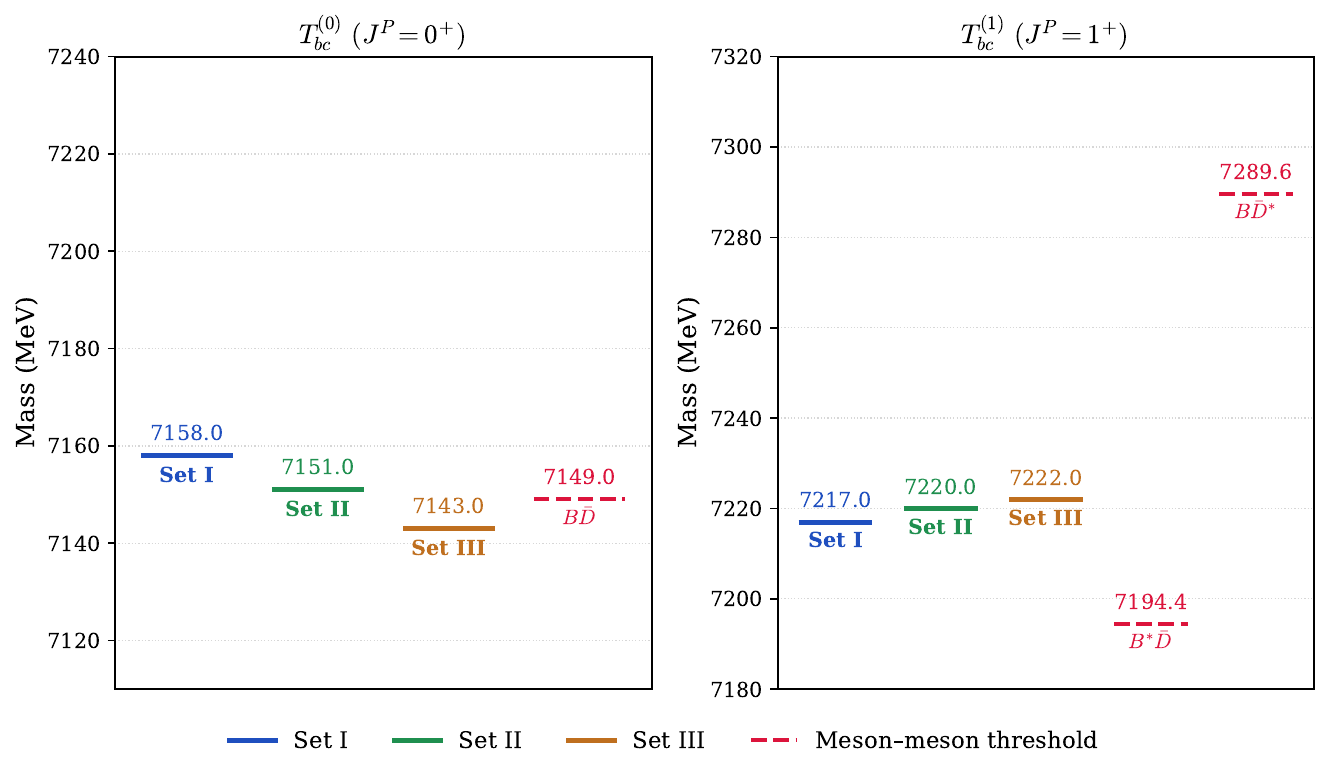}
\caption{Predicted masses of the doubly-heavy tetraquarks
$T_{bc}^{(0)}$ ($J^{P}=0^{+}$, left panel) and $T_{bc}^{(1)}$
($J^{P}=1^{+}$, right panel), obtained with the three parameter sets
employed in this work (solid lines, color-coded: Set~I blue, Set~II
green, Set~III orange), compared with the lowest open-flavor
meson--meson thresholds carrying the same quantum numbers (red dashed
lines). All masses are given in MeV.}
\label{fig:Tbc_levels}
\end{figure*}

The lowest meson--meson thresholds carrying the same flavor and
$J^{P}$ quantum numbers as the $T_{bc}$ system are, using
PDG-averaged meson masses \cite{ParticleDataGroup:2024cfk},
\begin{align}
B\bar D \;&:\; 5279.34 + 1869.66 \;=\; 7149.00~\text{MeV}
            \quad (0^{+}), \label{eq:thr_first}\\
B^{*}\bar D \;&:\; 5324.71 + 1869.66 \;=\; 7194.37~\text{MeV}
            \quad (1^{+}), \\
B\bar D^{*} \;&:\; 5279.34 + 2010.26 \;=\; 7289.60~\text{MeV}
            \quad (1^{+}), \\
B^{*}\bar D^{*} \;&:\; 5324.71 + 2010.26 \;=\; 7334.97~\text{MeV}.
            \label{eq:thr_last}
\end{align}
The last of these, which carries $J^{P}=0^{+},1^{+},2^{+}$, plays no
role for the isoscalar ground states discussed in this subsection,
since they lie far below it; it is, however, the relevant reference
for the isovector members of the multiplet considered in
Sec.~\ref{sec:other}.
A compact visual summary of the resulting spectroscopic situation is
provided in Fig.~\ref{fig:Tbc_levels}, where the predicted masses are
displayed alongside these thresholds.

The figure makes manifest two qualitatively distinct dynamical
situations. In the scalar sector (left panel), the three central
values bracket the $B\bar D$ threshold at $7149.0$~MeV with
displacements of $+9$, $+2$, and $-6$~MeV for Sets~I, II, and III,
respectively. These displacements are far smaller than the
$\mathcal{O}(60~\text{MeV})$ parametric uncertainty on the absolute
masses established above, so the statement supported by the
calculation is that the scalar eigenvalue is compatible with the
$B\bar D$ threshold, not that its position relative to threshold is
determined: at the central values of the inputs, $T_{bc}^{(0)}$ is
effectively sitting on the two-meson threshold, and the question of
whether the physical state would emerge as a genuinely bound state or
as a narrow near-threshold structure cannot be decided within the
present single-channel framework.
Should $M(T_{bc}^{(0)}) < M_B + M_{\bar D}$, as Set~III already
suggests, $S$-wave hadronic decay into $B\bar D$ is kinematically
forbidden, leaving only the weak $b\to c\ell\bar\nu$ and
$b\to c\bar ud$ transitions, the radiative
$T_{bc}^{(0)}\to B\bar D\gamma$ channel, and possible three-body
modes through the off-shell tails of $B^{*}$ and $\bar D^{*}$. For
the analogous $T_{cc}^{+}(3875)$ system, the latter mechanism is in
fact the dominant decay channel, with $T_{cc}^{+}\to D^{0}D^{0}\pi^{+}$
proceeding through the off-shell tail of the $D^{*+}$. For
$T_{bc}^{(0)}$ the corresponding $B^{*}$ and $\bar D^{*}$ tails lie
$\sim 45$ and $\sim 140$~MeV above the elastic threshold,
respectively, and are therefore kinematically more suppressed than
the $D^{*+}$ tail in the charm system; if the resulting radiative
and three-body modes could be neglected, the residual weak width
$\Gamma\sim \mathcal{O}(10^{-10})$~MeV would constitute a distinctive
signature of a strong-interaction-stable tetraquark. We stress that
this scenario is conditional on $M(T_{bc}^{(0)})$ lying below
threshold, which the present calculation cannot establish, and that a
dedicated estimate of the radiative and three-body channels would be
needed before any quantitative statement about the width can be made. The
near-tangency visible in Fig.~\ref{fig:Tbc_levels} thus strongly
motivates a dedicated coupled-channel analysis incorporating the
$B\bar D$ continuum explicitly, since the physical pole may well
migrate from the real axis onto the second Riemann sheet depending on
the residual $B\bar D$ interaction.

In the axial-vector sector (right panel), by contrast, the three
central values form a tight cluster around $7220$~MeV, positioned
above $B^{*}\bar D$ (by $23$--$28$~MeV) and well
below $B\bar D^{*}$ (by $\sim 70$~MeV) for all parameter sets. Within
the single-channel framework this locates $T_{bc}^{(1)}$ as a natural
candidate for an $S$-wave resonance in the $B^{*}\bar D$ channel,
with the caveat, spelled out in Sec.~\ref{sec:hamiltonian}, that an
eigenvalue above a two-meson threshold does not by itself establish a
resonance: whether the corresponding pole of the scattering amplitude
survives as a resonance, and at what position and width, can only be
determined from the amplitude itself. Should such a resonance exist
near the eigenvalue found here, the $\sim 25$~MeV excess above
threshold would be large enough to open the $B^{*}\bar D$ decay
channel but small enough that phase-space suppression and the
proximity of the $B^{*}\bar D$ cusp would visibly distort the
invariant-mass line shape---a situation reminiscent of
$T_{cc}^{+}(3875)$ relative to $D^{*+}D^{0}$---so that a Flatt\'e-type
or explicitly coupled-channel parametrization would be needed in the
experimental analysis.

The expected size of the neglected coupled-channel effects can be
estimated on general grounds. The $\delta$--$\bar\delta'$
configuration couples to the open and nearby closed
$B^{(*)}\bar D^{(*)}$ channels through light-quark rearrangement, and
this coupling generates a self-energy whose real part shifts the bare
eigenvalue. For a compact state coupled in $S$-wave to a two-meson
continuum, the magnitude of this shift is set by the same coupling
that controls the hadronic width of an above-threshold state and the
level repulsion from a below-threshold channel; for hadronic
couplings of typical strength and the near-threshold momenta relevant
here ($k \lesssim 300$~MeV), both dimensional analysis and explicit
coupled-channel studies of analogous doubly-heavy systems---including
the $T_{cc}^{+}(3875)$, for which the compact-core pole is drawn
toward the $D^{*+}D^{0}$ threshold by the channel coupling---indicate
shifts of order $10$--$30$~MeV, predominantly attractive, i.e.\
directed toward or below the nearest threshold. Three consequences
follow. First, for the scalar state, a downward shift of this size
would move an eigenvalue lying a few MeV above $B\bar D$ below
threshold, reinforcing the conclusion that the bound-versus-unbound
character of $T_{bc}^{(0)}$ cannot be decided at the present level of
the analysis. Second, for the axial-vector state, an attractive shift
of $10$--$30$~MeV would move the pole from $\sim 25$~MeV above
$B^{*}\bar D$ into the immediate vicinity of that threshold,
substantially reducing---and possibly closing---the gap to the
lattice determinations discussed below. Third, since these shifts are
comparable to the displacements from threshold on which the
qualitative classification of the states rests, the coupled-channel
dynamics is not a small correction in this system, and the
classifications offered here must be regarded as provisional until a
calculation incorporating the $B^{(*)}\bar D^{(*)}$ continua
explicitly becomes available.

The contrast between the two channels---a quasi-bound scalar versus a
near-threshold resonant axial-vector---is a direct consequence of the
interplay between the $60$--$80$~MeV hyperfine splitting found here
and the heavy-quark-symmetry-breaking pattern of the surrounding
open-flavor spectrum. Because the $D^{*}$--$D$ hyperfine gap
($\sim 140$~MeV) is substantially larger than the $B^{*}$--$B$ gap
($\sim 45$~MeV), exciting the charmed meson costs markedly more
energy than exciting the bottom meson; this pushes $B\bar D^{*}$ far
above $B^{*}\bar D$ and opens a wide window in which $T_{bc}^{(1)}$
can be trapped as a narrow resonance. Fig.~\ref{fig:Tbc_levels} thereby
encapsulates, in a single glance, both the internal tetraquark
dynamics and the external heavy-quark-symmetry-breaking pattern that
shapes the accessible decay channels. Any future observation that
inverts this hierarchy would point to physics beyond the
diquark--antidiquark picture adopted in this work.

Our results can be confronted with a growing body of theoretical
estimates obtained from a variety of independent methods.
In QCD sum-rule analyses, Ref.~\cite{Chen:2013aba} finds
the $T_{bc}$ states in the range
$7.0$--$7.2$~GeV. Constituent quark-model calculations, including
the relativistic diquark-antidiquark treatment of
Ref.~\cite{Ebert:2007rn} and the chromomagnetic-interaction
analyses of Refs.~\cite{Karliner:2017qjm,Eichten:2017ffp},
cluster around $7.1$--$7.3$~GeV, with
Ref.~\cite{Karliner:2017qjm} noting that the $I(J^P)=0(1^+)$
$T_{bc}$ state lies close to but possibly above the
$B^*\bar{D}$ threshold. A similar conclusion is reached in
Ref.~\cite{Park:2018wjk} within a constituent quark model with
parameters fitted to the $\Xi_{cc}^{++}$ mass. More recently,
the improved chromomagnetic-interaction model of
Ref.~\cite{Guo:2021yws} and the quark-model study of
Ref.~\cite{Carames:2018tpe} find $T_{bc}$
ground-state masses in the range $7.15$--$7.25$~GeV. Our values
fall comfortably within this band, while the hyperfine splitting
$\Delta_{\rm HF}\sim 60$--$80$~MeV is somewhat softer than the
$\sim 100$~MeV typical of pure chromomagnetic estimates---a
difference that traces back to the inclusion of the spatial
wave-function dependence of the spin-spin operator and to the
moderate values of $\kappa_{\bar{b}\bar{c}}$ adopted in the
present work.

A more demanding benchmark is provided by the recent lattice-QCD
results, which we now compare with systematically. Table~\ref{tab:lattice} summarizes, for each available lattice
calculation of the isoscalar $T_{bc}$ system, the interpolating-operator basis, the channels studied, and the quantitative outcome,
alongside the eigenvalues obtained in this work. The table makes explicit that the residual differences among the lattice studies
themselves---bound by tens of MeV in Ref.~\cite{Padmanath:2023rdu} versus at most a few MeV in Ref.~\cite{Alexandrou:2023cqg}---are
correlated with the operator content (local versus nonlocal scattering operators), the treatment of the left-hand cut from one-pion
exchange, and the chiral and continuum extrapolations, and are therefore of the same nature as the systematic effects that separate
those results from the present constituent calculation.

\begin{table*}[t]
\centering
\caption{Comparison of the present results with lattice-QCD determinations of the isoscalar $T_{bc}$ system. For each study we list the main methodological ingredients, the channels analyzed, and the quantitative outcome. Binding energies are quoted relative to the lowest relevant threshold ($B\bar D$ for $0^+$, $B^{*}\bar D$ for $1^+$). The entries of this work are single-channel eigenvalues with the parametric uncertainty discussed in the text, not pole positions.}
\label{tab:lattice}
\renewcommand{\arraystretch}{1.25}
\begin{tabular}{llll} \hline
Reference & Method and operator basis & Channels & Result \\ \hline
Ref.~\cite{Francis:2018jyb} & NRQCD $b$, relativistic $c$; local operators; & $0(1^+)$ & bound, $-61 < \Delta E < -15$~MeV \\
 & $m_\pi \gtrsim 164$~MeV, $m_\pi L = 2.4$ on lightest ensemble & & \\
Ref.~\cite{Hudspith:2020tdf} & box-sink construction, enlarged basis; & $0(0^+)$, $0(1^+)$ & no binding; $1^+$ marginally above $B^{*}\bar D$ \\
 & $m_\pi \simeq 192$~MeV, $m_\pi L = 4.2$ & & \\
Ref.~\cite{Meinel:2022lzo} & domain-wall light, NRQCD $b$, clover $c$; & $0(0^+)$, $0(1^+)$ & consistent with thresholds; shallow bound \\
 & nonlocal meson--meson sinks; physical $m_\pi$ included & & state not excluded \\
Ref.~\cite{Padmanath:2023rdu} & HISQ; local diquark--antidiquark $+$ two-meson basis; & $0(1^+)$ & bound, $-43(^{+6}_{-7})(^{+14}_{-24})$~MeV; \\
 & L\"uscher analysis; continuum and chiral extrapolation & & unbound above $M^{*}_{\rm ps}=2.73(21)(19)$~GeV \\
Ref.~\cite{Alexandrou:2023cqg} & $B^{(*)}\bar D$ scattering operators at source and sink; & $0(0^+)$, $0(1^+)$ & shallow (possibly virtual) poles a few MeV \\
 & effective-range analysis; $m_\pi \simeq 220$~MeV & & below threshold; hints of broad resonances \\
This work & single-channel BO dynamical diquark model; & $0(0^+)$, $0(1^+)$ & $0^+$: $-6$ to $+9$~MeV about $B\bar D$; \\
 & $\bar{\mathbf{3}}_c \otimes \mathbf{3}_c$; eigenvalues, not pole positions & & $1^+$: $+23$ to $+28$~MeV above $B^{*}\bar D$ \\ \hline
\end{tabular}
\end{table*}

The two-meson-plus-diquark variational analysis of
Ref.~\cite{Padmanath:2023rdu} reports a binding energy of
$-43(^{+6}_{-7})(^{+14}_{-24})$~MeV below $B^{*}\bar{D}$ for the
$I(J^P)=0(1^+)$ ground state, which corresponds to a physical mass
of $\simeq 7151$~MeV when added to the experimental
$B^{*}\bar{D}$ threshold of $7194.4$~MeV. The
$B^{(*)}\bar{D}$-scattering analysis of
Ref.~\cite{Alexandrou:2023cqg}, by contrast, locates the
near-threshold pole at most a few MeV below $B^{*}\bar{D}$, i.e.\
near $7190$~MeV. Our axial-vector eigenvalue
$M(T_{bc}^{(1)})\simeq 7220$~MeV thus lies $\sim 70$~MeV above the
binding-energy lattice value of Ref.~\cite{Padmanath:2023rdu} and
$\sim 30$~MeV above the scattering-pole value of
Ref.~\cite{Alexandrou:2023cqg}, and on the wrong side of
$B^{*}\bar{D}$ in both cases. This $\mathcal{O}(30$--$70)$~MeV gap
lies within the systematic uncertainty generically expected of a
constituent BO description, in which the dominant systematic
of order $\Lambda_{\mathrm{QCD}}$ affects primarily fine-structure
quantities of the kind that determine the position of the pole
relative to threshold; we view our result as the constituent
$\delta$--$\bar{\delta}'$ value, and a comparison with lattice at
better than $\mathcal{O}(\Lambda_{\mathrm{QCD}})$ accuracy lies
beyond the scope of the present analysis. We note in this
connection that the single-channel $\Sigma_g^+(1S)$ description
adopted here omits two physical effects that are explicitly
included in the lattice analyses and that are natural candidates
for explaining the residual gap: long-range one-pion exchange
between the $B^{(*)}$ and $\bar D^{(*)}$ clusters, which generates
a left-hand cut in the elastic amplitude, and explicit
coupled-channel $B^{(*)}\bar D^{(*)}$ dynamics, which can
generically pull the physical pole downward toward---or below---the
nearest two-meson threshold. A quantitative incorporation of
either ingredient lies beyond the present single-channel BO
treatment.

From an experimental standpoint, a resonant
structure in the $B\bar D$ invariant-mass distribution near
$7.15$~GeV, displaced from threshold by at most a handful of MeV,
would be a natural candidate for $T_{bc}^{(0)}$, while a
structure in $B^{*}\bar D$ around $7.22$~GeV, roughly $25$~MeV
above threshold, would be the expected signature of $T_{bc}^{(1)}$.
The absence of a nearby eigenvalue in the $B\bar D^{*}$ channel
further suggests that the axial-vector state, if observed, could
display a pronounced branching asymmetry in favor of $B^{*}\bar D$
over $B\bar D^{*}$, which would provide an additional discriminator
against competing molecular interpretations; a quantitative statement
would, however, require explicit decay-width calculations that are
not performed here. With the high-luminosity runs
of LHCb and the upcoming Belle~II dataset, $bc\bar q\bar q$ states
in the $7.1$--$7.3$~GeV window are within reach, and a
confirmation or refutation of the hierarchy and near-threshold
character suggested here would significantly sharpen our
understanding of multiquark dynamics in the doubly-heavy sector.

\subsection{Other configurations: the complete $\Sigma_g^+(1S)$
multiplet and the color-sextet channel}
\label{sec:other}

The two isoscalar states discussed so far exhaust neither the
$\Sigma_g^+(1S)$ multiplet nor the color structures available to a
$bc\bar q\bar q$ system. In this subsection we complete the
spectrum of the multiplet, and we estimate the effect of the
$\bar{\mathbf{6}}\otimes\mathbf{6}$ configuration that was set aside
in Sec.~\ref{sec:ddm}.

\subsubsection{The isovector sector}

The four isovector members of the multiplet follow from
Eqs.~\eqref{eq:M_Ttilde} and \eqref{eq:M_Ttildeprime} without any
additional numerical input. Comparing them with
Eqs.~\eqref{eq:M_T0}--\eqref{eq:M_T1} one finds the exact relations
\begin{align}
M\!\left(\widetilde{T}_{bc}^{(J)}\right)
  &= M\!\left(T_{bc}^{(1)}\right) + 2\kappa_{ud},
  \qquad J=0,1,2, \label{eq:shift1}\\
M\!\left(\widetilde{T}_{bc}^{\prime(1)}\right)
  &= M\!\left(T_{bc}^{(0)}\right) + 2\kappa_{ud},
  \label{eq:shift2}
\end{align}
so that the entire isovector sector is obtained from the isoscalar
one by a rigid upward shift of $2\kappa_{ud} \approx 206$~MeV, the
energy cost of promoting the light $[ud]$ pair from the good to the
bad diquark configuration. This is a structural prediction of the
isospin--spin lock of Eq.~\eqref{eq:IsospinSpinLock}: the same
operator that generates the good--bad diquark splitting also fixes
the isoscalar--isovector gap, without any isospin-dependent
interaction being introduced by hand. The resulting spectrum of the
full multiplet is collected in Table~\ref{tab:multiplet}.

\begin{table*}[t]
\centering
\caption{The complete $\Sigma_g^+(1S)$ multiplet of the $T_{bc}$
system. The isovector masses follow from
Eqs.~\eqref{eq:shift1}--\eqref{eq:shift2} and require no additional
input. The last column gives the position relative to the highest
$S$-wave threshold with the same quantum numbers ($B\bar D$ for the
isoscalar $0^{+}$, $B^{*}\bar D$ for the isoscalar $1^{+}$, and
$B^{*}\bar D^{*}=7335.0$~MeV for all isovector states). Masses in
GeV.}
\label{tab:multiplet}
\renewcommand{\arraystretch}{1.2}
\small
\begin{tabular}{llccccc} \hline
State & $I(J^{P})$ & Configuration & Set~I & Set~II & Set~III & Position \\ \hline
$T_{bc}^{(0)}$ & $0(0^{+})$ & $|[ud],[\bar b\bar c]\rangle_{0}$ & 7.158 & 7.151 & 7.143 & $\pm 10$ MeV of $B\bar D$ \\
$T_{bc}^{(1)}$ & $0(1^{+})$ & $|[ud],\{\bar b\bar c\}\rangle_{1}$ & 7.217 & 7.220 & 7.222 & $+23$ to $+28$ MeV \\
$\widetilde{T}_{bc}^{\prime(1)}$ & $1(1^{+})$ & $|\{ud\},[\bar b\bar c]\rangle_{1}$ & 7.364 & 7.357 & 7.349 & $+14$ to $+29$ MeV \\
$\widetilde{T}_{bc}^{(0)}$ & $1(0^{+})$ & $|\{ud\},\{\bar b\bar c\}\rangle_{0}$ & 7.423 & 7.426 & 7.428 & $+88$ to $+93$ MeV \\
$\widetilde{T}_{bc}^{(1)}$ & $1(1^{+})$ & $|\{ud\},\{\bar b\bar c\}\rangle_{1}$ & 7.423 & 7.426 & 7.428 & $+88$ to $+93$ MeV \\
$\widetilde{T}_{bc}^{(2)}$ & $1(2^{+})$ & $|\{ud\},\{\bar b\bar c\}\rangle_{2}$ & 7.423 & 7.426 & 7.428 & $+88$ to $+93$ MeV \\ \hline
\end{tabular}
\end{table*}

Three features of Table~\ref{tab:multiplet} deserve comment. First,
every isovector state lies above \emph{all four} open-flavor
thresholds of Eqs.~(\ref{eq:thr_first}--\ref{eq:thr_last}), including
$B^{*}\bar D^{*}$. All $S$-wave decay channels allowed by the
quantum numbers are therefore open, with sizable phase space, and
these states are expected to be broad. This is consistent with the
absence of any lattice-QCD signal for isovector $T_{bc}$ states, and
it explains why the isoscalar channels---the only ones in which the
light pair can occupy the good-diquark configuration---are the ones
in which a narrow, potentially strong-interaction-stable structure
can be expected. Second, the three states
$\widetilde{T}_{bc}^{(0,1,2)}$ are exactly degenerate at the level of
$H_\kappa$, as explained in Sec.~\ref{sec:hamiltonian}; a physical
splitting of this triplet requires spin-orbit and tensor forces,
which vanish identically for $L=0$ in the Hamiltonian adopted here
and would first be generated by the mixing with excited BO potentials
and by relativistic corrections. The degeneracy pattern of
Table~\ref{tab:multiplet} is therefore a prediction of the leading
BO approximation and not a statement about the physical spectrum at
the $10$~MeV level. Third, since $\widetilde{T}_{bc}^{\prime(1)}$ and
$\widetilde{T}_{bc}^{(1)}$ carry identical conserved quantum numbers,
the physical $I=1$, $J^P=1^+$ eigenstates will be mixtures of the two
internal spin configurations once interactions beyond $H_\kappa$ are
included; the entries of Table~\ref{tab:multiplet} for these two
states are to be read accordingly.

Beyond the $\Sigma_g^+(1S)$ multiplet, the same framework supports a
tower of radial and orbital excitations. The first radially excited
$S$-wave level of the $\Sigma_g^+$ potential lies roughly $0.6$~GeV
above the ground state for the reduced masses relevant here, and the
excited BO potentials $\Pi_u$ and $\Sigma_u^-$ generate hybrid-like
multiplets in the same region, so that the $7.5$--$8.0$~GeV window is
expected to be densely populated. Since all of these states lie far
above the open-flavor thresholds, and since the corresponding
level orderings are considerably more sensitive to the BO potentials
and to the neglected spin-dependent forces than the ground-state
doublet, we do not attempt a quantitative treatment here.

\subsubsection{The $\bar{\mathbf{6}}\otimes\mathbf{6}$ configuration}

As emphasized in Sec.~\ref{sec:ddm}, the argument that discards the
sextet channel on the grounds that the interaction between two
(anti)quarks in the $\mathbf{6}_c$ ($\bar{\mathbf{6}}_c$)
representation is repulsive is incomplete, because the interaction
\emph{between} a $\mathbf{6}_c$ cluster and a $\bar{\mathbf{6}}_c$
cluster is attractive and, by Casimir scaling, stronger than the
$\bar{\mathbf{3}}_c$--$\mathbf{3}_c$ one. Both effects can be
assessed, at least semiquantitatively, within the present framework.

For the inter-cluster dynamics, Casimir scaling of the static energy
gives
\begin{equation}
V_{\bar{\mathbf{6}}\mathbf{6}}(r) - V_0
\;\simeq\; \frac{C_6}{C_3}\left[V_{\Sigma_g^+}(r) - V_0\right],
\qquad \frac{C_6}{C_3}=\frac{5}{2},
\label{eq:casimir}
\end{equation}
where $V_0$ denotes the additive self-energy constant, which is not
Casimir-scalable and which, as discussed in Sec.~\ref{sec:inputs}, is
in any case degenerate with the cluster masses. Lattice measurements
of static potentials in higher representations support
Eq.~\eqref{eq:casimir} at the few-percent level out to separations of
about $1$~fm~\cite{Bali:2000un}, comfortably covering the region
$\langle r\rangle\simeq 0.45$~fm relevant here. Solving
Eq.~\eqref{eq:schroedinger} with the scaled potential and the same
reduced mass, we find that the $S$-wave eigenvalue of the sextet
channel lies $0.1$--$0.3$~GeV \emph{above} the antitriplet one, the
spread reflecting the Coulomb-to-linear apportionment of the
parametrization: the factor $5/2$ deepens the short-distance well but
also steepens the confining rise, and at the separations relevant
here the latter effect dominates. The associated wave function is
correspondingly more compact, $\langle r\rangle \simeq 0.3$~fm. To
this must be added the intra-cluster penalty. One-gluon exchange
between two quarks in the $\mathbf{6}_c$ channel carries a color
factor equal to $-1/2$ times the $\bar{\mathbf{3}}_c$ one, so that
the interaction is repulsive and no compact $\mathbf{6}_c$ cluster is
formed; strictly speaking, the cluster description itself ceases to
apply, and the energy of the configuration is raised further with
respect to the estimate above.

Taken together, these considerations place the diagonal energy of the
$\bar{\mathbf{6}}\otimes\mathbf{6}$ configuration several hundred MeV
above the $\bar{\mathbf{3}}\otimes\mathbf{3}$ one. Its effect on the
ground state is then of second order,
\begin{align}
\delta E \;&\simeq\; -\frac{|H_{\bar{3}3,\bar{6}6}|^{2}}{\Delta E_{36}}
\nonumber\\
&\approx\; -10~\text{MeV}
\left(\frac{|H_{\bar{3}3,\bar{6}6}|}{100~\text{MeV}}\right)^{2}
\left(\frac{250~\text{MeV}}{\Delta E_{36}}\right),
\label{eq:mixing}
\end{align}
i.e.\ a downward shift of order $10$~MeV, reaching a few tens of MeV
only for off-diagonal matrix elements at the upper end of what is
plausible for a color-rearrangement interaction. Three consequences
follow. The sextet channel is not a competing leading configuration,
so the restriction to $\bar{\mathbf{3}}\otimes\mathbf{3}$ adopted
here is justified as a leading approximation; the neglected mixing
acts, by the variational argument of Sec.~\ref{sec:ddm}, in the
direction of additional binding, so that the eigenvalues of
Tables~\ref{tab:Tbc0}--\ref{tab:Tbc1} remain upper limits within the
model; and the magnitude of the induced shift, $\mathcal{O}(10)$~MeV,
is comparable to that of the neglected coupled-channel effects
estimated in Sec.~\ref{sec:thresholds} and is already contained in
the uncertainty budget of Sec.~\ref{sec:inputs}. It should be
stressed that Eqs.~\eqref{eq:casimir} and \eqref{eq:mixing} are
estimates within a static, two-cluster picture; a genuine assessment
requires a four-body calculation retaining the full color space, of
the kind performed in Refs.~\cite{Carames:2018tpe,Deng:2018kly,Yang:2019itm,Tan:2020ldi},
where sizable components beyond a single
$\bar{\mathbf{3}}\otimes\mathbf{3}$ configuration are indeed found.

\section{Conclusions}
\label{sec:conclusions}

In this work we have carried out a systematic study of the
doubly-heavy open-flavor tetraquark states $T_{bc}^{(0)}$
($J^{P}=0^{+}$) and $T_{bc}^{(1)}$ ($J^{P}=1^{+}$) within the
dynamical diquark model, treating the $T_{bc}$ system as a
heavy antidiquark--light diquark pair propagating in the
$\Sigma_g^+(1S)$ BO potential extracted from
lattice QCD. The light-diquark mass, the heavy-antidiquark masses
of both spin configurations, and the light-diquark spin--spin
coupling were fixed from independent phenomenological inputs, while
the heavy-antidiquark chromomagnetic coupling
$\kappa_{\bar{b}\bar{c}}$ was varied over the range $10$--$20$~MeV,
in order to probe the sensitivity of the spectrum to the least
constrained parameter of the model. A full parametric uncertainty
budget for all inputs was presented in Sec.~\ref{sec:inputs},
together with an explicit discussion of why the inputs necessarily
originate from different frameworks, of the degeneracy between the
constituent cluster masses and the additive constant of the static
potential, and of the two limiting conventions for apportioning the
heavy-antidiquark spin dependence, which bracket the hyperfine
splitting between $20$ and $80$~MeV. We also completed the
$\Sigma_g^+(1S)$ multiplet by computing its isovector members, and
estimated the size of the neglected color mixing, non-adiabatic
corrections, and coupled-channel dynamics.

The principal numerical findings can be summarized as follows. The
scalar eigenvalue lies in the narrow window
$M(T_{bc}^{(0)}) = 7.143$--$7.158$~GeV and the axial-vector
eigenvalue in $M(T_{bc}^{(1)}) = 7.217$--$7.222$~GeV, with
$\kappa_{\bar b\bar c}$-induced spreads of only $\sim 15$ and
$\sim 5$~MeV, respectively; the overall parametric uncertainty on the
absolute masses is of order $\pm 60$--$70$~MeV, dominated by the
light-diquark mass, but it is largely common to the two channels and
cancels in mass differences. The associated
hyperfine splitting evolves linearly with $\kappa_{\bar{b}\bar{c}}$
in the range $\Delta_{HF}\simeq 59$--$79$~MeV, the bulk of which
$(\sim 40~\text{MeV})$ comes from the kinematic mass gap between
the symmetric and antisymmetric heavy-antidiquark configurations,
while the slope $\partial\Delta_{HF}/\partial\kappa_{\bar{b}\bar{c}}=2$
follows analytically from the
$\langle\vec{S}_{\bar b}\!\cdot\!\vec{S}_{\bar c}\rangle$
eigenvalues of the two heavy-antidiquark configurations. Once the
heavy-antidiquark mass gap has been independently fixed, this linear
response endows the framework with genuine predictive power: a
future measurement of $\Delta_{HF}$ will then fix
$\kappa_{\bar{b}\bar{c}}$ in a regime so far inaccessible to either
lattice QCD or phenomenology. The spatial observables, the mean inter-cluster separation
$\langle r\rangle\simeq 0.45$--$0.46$~fm and the inverse mean radius
$\langle 1/r\rangle^{-1}\simeq 0.33$--$0.34$~fm, were found to be
essentially independent both of $\kappa_{\bar{b}\bar{c}}$, which
enters the radial equation only as an additive constant, and of
$J^{P}$: the $40$~MeV gap between the symmetric and antisymmetric
heavy-antidiquark configurations changes the reduced mass by only
$5\times 10^{-4}$, so that the two channels share the same radial
profile far below the precision at which the radii are quoted.
This indicates that the spatial wave function is governed almost
entirely by the heavy-antidiquark--light-diquark reduced mass and
that the chromomagnetic interaction acts as a small perturbation on
an otherwise nearly universal radial profile. The
ratio $\langle 1/r\rangle^{-1}/\langle r\rangle\simeq 0.73$--$0.74$
situates the $\Sigma_g^+(1S)$ wave function squarely between the
sharply localized and long-tailed limits, indicating that both the
short-range gluon exchange and the long-range confining string
contribute appreciably to the binding---a pattern familiar from heavy
quarkonia and now seen to extend to genuinely four-quark
configurations.

Confronting these results with the relevant open-flavor
thresholds reveals two qualitatively different dynamical
situations. The central value of the scalar $T_{bc}^{(0)}$ mass lies
essentially on the $B\bar D$ threshold, with displacements ranging
from $+9$~MeV (Set~I) to $-6$~MeV (Set~III). Whether the physical
state would ultimately emerge as a genuinely bound, weakly decaying
tetraquark---in close analogy with the role played by the
$D^{*+}D^{0}$ threshold for $T_{cc}^{+}(3875)$---or as a narrow
near-threshold structure cannot be decided within the present
framework: the answer depends not only on the precise value of
$\kappa_{\bar{b}\bar{c}}$ but also on the parametric uncertainties of
the inputs and on the neglected coupled-channel dynamics, whose
estimated $\mathcal{O}(10$--$30)$~MeV effects are comparable to the
displacement from threshold, and a definitive answer will require a
coupled-channel analysis incorporating the $B\bar D$ continuum
explicitly so as to track the migration of the physical pole
between the first and second Riemann sheets. The axial-vector
$T_{bc}^{(1)}$ eigenvalue, in contrast, is consistently found
$23$--$28$~MeV above $B^{*}\bar D$ but $\sim 70$~MeV below
$B\bar D^{*}$, suggesting an interpretation as an $S$-wave
resonance candidate in the $B^{*}\bar D$ channel, with a line shape
that would be strongly distorted by the nearby cusp; here too, the
identification of the eigenvalue with a physical pole position is
subject to the coupled-channel caveats discussed in
Sec.~\ref{sec:thresholds}. The
contrast between the two channels---a quasi-bound scalar versus a
near-threshold resonant axial-vector---is not an accident of the
model but a direct consequence of the interplay between the
$\sim 70$~MeV tetraquark hyperfine splitting and the
heavy-quark-symmetry-breaking pattern of the surrounding
open-flavor spectrum, in which the $\sim 140$~MeV $D^{*}$--$D$ gap
substantially exceeds the $\sim 45$~MeV $B^{*}$--$B$ gap.

Completing the $\Sigma_g^+(1S)$ multiplet, we find that its four
isovector members follow from the isoscalar ones by a rigid shift of
$2\kappa_{ud}\approx 206$~MeV, a structural consequence of the
isospin--spin lock rather than of an independent isospin-dependent
force, and that all of them lie above every $B^{(*)}\bar D^{(*)}$
threshold, so that they are expected to be broad and are unlikely to
be observed as distinct structures. We have also revisited the
restriction to the $\bar{\mathbf{3}}\otimes\mathbf{3}$ color
configuration. The interaction between a $\mathbf{6}_c$ cluster and
a $\bar{\mathbf{6}}_c$ cluster is attractive and, by Casimir
scaling, stronger than the $\bar{\mathbf{3}}_c$--$\mathbf{3}_c$ one
by the factor $5/2$, so that the sextet channel cannot be dismissed
on the basis of the intra-cluster repulsion alone; solving the radial
problem in the Casimir-scaled potential, however, places the sextet
configuration $0.1$--$0.3$~GeV higher, because the steeper confining
rise outweighs the deeper short-distance well at the separations
relevant here, and the intra-cluster repulsion raises it further. The
resulting second-order admixture shifts the ground state downward by
$\mathcal{O}(10)$~MeV, comparable to the neglected coupled-channel
effects and already covered by the quoted systematic uncertainty.

From a broader perspective, the present analysis illustrates how
the dynamical diquark framework, supplemented with lattice-QCD
BO potentials and a minimal set of independently
constrained inputs, can deliver definite and falsifiable
central-value estimates for exotic states in the doubly-heavy sector
with very modest parametric freedom, provided the systematic
limitations spelled out above are kept in view. Several natural extensions invite further
investigation. The inclusion of explicit coupled-channel
$B^{(*)}\bar D^{(*)}$ dynamics is essential to convert the
near-threshold scalar result into a quantitative statement
about binding versus virtuality. The tower of orbitally and
radially excited states built on higher BO
potentials---most notably $\Sigma_u^-$ and $\Pi_u$, which generate
hybrid-like multiplets---would extend the spectrum to the
$7.5$--$8.0$~GeV region and provide additional tests of the
diquark organization. Electromagnetic and weak transition rates,
together with production cross sections in $pp$ collisions and
$B_c$ decays, would supply the dynamical observables needed to
discriminate the compact tetraquark interpretation from molecular
or hadrocharmonium alternatives. Last, an analogous treatment of
the $T_{bb}$ and $T_{cc}$ sectors,
using the same BO potential and the corresponding
heavy-antidiquark masses, would test the universality of the
underlying picture and clarify the systematic trends predicted by
heavy-quark symmetry.

On the experimental side, $T_{bc}$ states in the
$7.1$--$7.3$~GeV window are within reach of the high-luminosity
runs of LHCb and the upcoming Belle~II dataset. A resonant
structure observed in the $B\bar D$ invariant-mass distribution
near $7.15$~GeV, displaced from threshold by no more than a
handful of MeV, would be a natural candidate for $T_{bc}^{(0)}$,
while a structure in $B^{*}\bar D$ around $7.22$~GeV, roughly
$25$~MeV above threshold, would constitute the expected signature
of $T_{bc}^{(1)}$. A pronounced branching asymmetry in favor of
$B^{*}\bar D$ over $B\bar D^{*}$ would further reinforce the
compact-tetraquark interpretation against molecular alternatives.
The discovery---or non-observation---of these states with the
mass hierarchy and near-threshold character estimated here would
significantly sharpen our understanding of multiquark dynamics in
the doubly-heavy sector and, by pinning down
$\kappa_{\bar{b}\bar{c}}$, provide a long-sought empirical anchor
for the chromomagnetic interaction between two distinct heavy
antiquarks.

\section*{Acknowledgment}
This work is supported by Scientific Research Projects Coordination Unit of Ondokuz Mayis University with project number BAP01-2025-5761.

\bibliography{bcudtetraquark}

\begin{thebibliography}{10}
\expandafter\ifx\csname url\endcsname\relax
  \def\url#1{\texttt{#1}}\fi
\expandafter\ifx\csname urlprefix\endcsname\relax\def\urlprefix{URL }\fi
\expandafter\ifx\csname href\endcsname\relax
  \def\href#1#2{#2} \def\path#1{#1}\fi

\bibitem{LHCb:2021vvq}
R.~Aaij, et~al., {Observation of an exotic narrow doubly charmed tetraquark},
  Nature Phys. 18~(7) (2022) 751--754.
\newblock \href {http://arxiv.org/abs/2109.01038} {\path{arXiv:2109.01038}},
  \href {https://doi.org/10.1038/s41567-022-01614-y}
  {\path{doi:10.1038/s41567-022-01614-y}}.

\bibitem{LHCb:2021auc}
R.~Aaij, et~al., {Study of the doubly charmed tetraquark $T_{cc}^+$}, Nature
  Commun. 13~(1) (2022) 3351.
\newblock \href {http://arxiv.org/abs/2109.01056} {\path{arXiv:2109.01056}},
  \href {https://doi.org/10.1038/s41467-022-30206-w}
  {\path{doi:10.1038/s41467-022-30206-w}}.

\bibitem{Ader:1981db}
J.~P. Ader, J.~M. Richard, P.~Taxil, {Do narrow heavy multi-quark states
  exist?}, Phys. Rev. D 25 (1982) 2370.
\newblock \href {https://doi.org/10.1103/PhysRevD.25.2370}
  {\path{doi:10.1103/PhysRevD.25.2370}}.

\bibitem{Carlson:1987hh}
J.~Carlson, L.~Heller, J.~A. Tjon, {Stability of Dimesons}, Phys. Rev. D 37
  (1988) 744.
\newblock \href {https://doi.org/10.1103/PhysRevD.37.744}
  {\path{doi:10.1103/PhysRevD.37.744}}.

\bibitem{Manohar:1992nd}
A.~V. Manohar, M.~B. Wise, {Exotic $QQ\bar q\bar q$ states in QCD}, Nucl. Phys.
  B 399 (1993) 17--33.
\newblock \href {http://arxiv.org/abs/hep-ph/9212236}
  {\path{arXiv:hep-ph/9212236}}, \href
  {https://doi.org/10.1016/0550-3213(93)90614-U}
  {\path{doi:10.1016/0550-3213(93)90614-U}}.

\bibitem{Eichten:2017ffp}
E.~J. Eichten, C.~Quigg, {Heavy-quark symmetry implies stable heavy tetraquark
  mesons $Q_iQ_j\bar q_k\bar q_l$}, Phys. Rev. Lett. 119~(20) (2017) 202002.
\newblock \href {http://arxiv.org/abs/1707.09575} {\path{arXiv:1707.09575}},
  \href {https://doi.org/10.1103/PhysRevLett.119.202002}
  {\path{doi:10.1103/PhysRevLett.119.202002}}.

\bibitem{Jaffe:2004ph}
R.~L. Jaffe, {Exotica}, Phys. Rept. 409 (2005) 1--45.
\newblock \href {http://arxiv.org/abs/hep-ph/0409065}
  {\path{arXiv:hep-ph/0409065}}, \href
  {https://doi.org/10.1016/j.physrep.2004.11.005}
  {\path{doi:10.1016/j.physrep.2004.11.005}}.

\bibitem{Francis:2016hui}
A.~Francis, R.~J. Hudspith, R.~Lewis, K.~Maltman, {Lattice Prediction for
  Deeply Bound Doubly Heavy Tetraquarks}, Phys. Rev. Lett. 118~(14) (2017)
  142001.
\newblock \href {http://arxiv.org/abs/1607.05214} {\path{arXiv:1607.05214}},
  \href {https://doi.org/10.1103/PhysRevLett.118.142001}
  {\path{doi:10.1103/PhysRevLett.118.142001}}.

\bibitem{Junnarkar:2018twb}
P.~Junnarkar, N.~Mathur, M.~Padmanath, {Study of doubly heavy tetraquarks in
  Lattice QCD}, Phys. Rev. D 99~(3) (2019) 034507.
\newblock \href {http://arxiv.org/abs/1810.12285} {\path{arXiv:1810.12285}},
  \href {https://doi.org/10.1103/PhysRevD.99.034507}
  {\path{doi:10.1103/PhysRevD.99.034507}}.

\bibitem{Leskovec:2019ioa}
L.~Leskovec, S.~Meinel, M.~Pflaumer, M.~Wagner, {Lattice QCD investigation of a
  doubly-bottom $\bar b\bar b ud$ tetraquark with quantum numbers
  $I(J^P)=0(1^+)$}, Phys. Rev. D 100~(1) (2019) 014503.
\newblock \href {http://arxiv.org/abs/1904.04197} {\path{arXiv:1904.04197}},
  \href {https://doi.org/10.1103/PhysRevD.100.014503}
  {\path{doi:10.1103/PhysRevD.100.014503}}.

\bibitem{Hudspith:2020tdf}
R.~J. Hudspith, B.~Colquhoun, A.~Francis, R.~Lewis, K.~Maltman, {A lattice
  investigation of exotic tetraquark channels}, Phys. Rev. D 102 (2020) 114506.
\newblock \href {http://arxiv.org/abs/2006.14294} {\path{arXiv:2006.14294}},
  \href {https://doi.org/10.1103/PhysRevD.102.114506}
  {\path{doi:10.1103/PhysRevD.102.114506}}.

\bibitem{Meinel:2022lzo}
S.~Meinel, M.~Pflaumer, M.~Wagner, {Search for $b\bar b\bar us$ and $b\bar
  c\bar ud$ tetraquark bound states using lattice QCD}, Phys. Rev. D 106~(3)
  (2022) 034507.
\newblock \href {http://arxiv.org/abs/2205.13982} {\path{arXiv:2205.13982}},
  \href {https://doi.org/10.1103/PhysRevD.106.034507}
  {\path{doi:10.1103/PhysRevD.106.034507}}.

\bibitem{Ali:2018xfq}
A.~Ali, Q.~Qin, W.~Wang, {Discovery potential of stable and near-threshold
  doubly heavy tetraquarks at the LHC}, Phys. Lett. B 785 (2018) 605--609.
\newblock \href {http://arxiv.org/abs/1806.09288} {\path{arXiv:1806.09288}},
  \href {https://doi.org/10.1016/j.physletb.2018.09.018}
  {\path{doi:10.1016/j.physletb.2018.09.018}}.

\bibitem{SilvestreBrac:1993ss}
B.~Silvestre-Brac, C.~Semay, {Systematics of $L=0$ $q^2\bar q^2$ systems}, Z.
  Phys. C 57 (1993) 273--282.
\newblock \href {https://doi.org/10.1007/BF01565058}
  {\path{doi:10.1007/BF01565058}}.

\bibitem{Semay:1994ht}
C.~Semay, B.~Silvestre-Brac, {Diquonia and potential models}, Z. Phys. C 61
  (1994) 271--275.
\newblock \href {https://doi.org/10.1007/BF01413104}
  {\path{doi:10.1007/BF01413104}}.

\bibitem{Ebert:2007rn}
D.~Ebert, R.~N. Faustov, V.~O. Galkin, W.~Lucha, {Masses of tetraquarks with
  two heavy quarks in the relativistic quark model}, Phys. Rev. D 76 (2007)
  114015.
\newblock \href {http://arxiv.org/abs/0706.3853} {\path{arXiv:0706.3853}},
  \href {https://doi.org/10.1103/PhysRevD.76.114015}
  {\path{doi:10.1103/PhysRevD.76.114015}}.

\bibitem{Park:2018wjk}
W.~Park, S.~Noh, S.~H. Lee, {Masses of the doubly heavy tetraquarks in a
  constituent quark model}, Nucl. Phys. A 983 (2019) 1--19.
\newblock \href {http://arxiv.org/abs/1809.05257} {\path{arXiv:1809.05257}},
  \href {https://doi.org/10.1016/j.nuclphysa.2018.12.019}
  {\path{doi:10.1016/j.nuclphysa.2018.12.019}}.

\bibitem{Carames:2018tpe}
T.~F. Caram{\'e}s, J.~Vijande, A.~Valcarce, {Exotic $bc\bar q\bar q$ four-quark
  states}, Phys. Rev. D 99~(1) (2019) 014006.
\newblock \href {http://arxiv.org/abs/1812.08991} {\path{arXiv:1812.08991}},
  \href {https://doi.org/10.1103/PhysRevD.99.014006}
  {\path{doi:10.1103/PhysRevD.99.014006}}.

\bibitem{Braaten:2020nwp}
E.~Braaten, L.-P. He, A.~Mohapatra, {Masses of doubly heavy tetraquarks with
  error bars}, Phys. Rev. D 103~(1) (2021) 016001.
\newblock \href {http://arxiv.org/abs/2006.08650} {\path{arXiv:2006.08650}},
  \href {https://doi.org/10.1103/PhysRevD.103.016001}
  {\path{doi:10.1103/PhysRevD.103.016001}}.

\bibitem{Lu:2020rog}
Q.-F. L{\"u}, D.-Y. Chen, Y.-B. Dong, {Masses of doubly heavy tetraquarks
  $T_{QQ'}$ in a relativized quark model}, Phys. Rev. D 102~(3) (2020) 034012.
\newblock \href {http://arxiv.org/abs/2006.08087} {\path{arXiv:2006.08087}},
  \href {https://doi.org/10.1103/PhysRevD.102.034012}
  {\path{doi:10.1103/PhysRevD.102.034012}}.

\bibitem{Deng:2018kly}
C.~Deng, H.~Chen, J.~Ping, {Systematical investigation on the stability of
  doubly heavy tetraquark states}, Eur. Phys. J. A 56~(1) (2020) 9.
\newblock \href {http://arxiv.org/abs/1811.06462} {\path{arXiv:1811.06462}},
  \href {https://doi.org/10.1140/epja/s10050-019-00012-y}
  {\path{doi:10.1140/epja/s10050-019-00012-y}}.

\bibitem{Yang:2019itm}
G.~Yang, J.~Ping, J.~Segovia, {Doubly-heavy tetraquarks}, Phys. Rev. D 101~(1)
  (2020) 014001.
\newblock \href {http://arxiv.org/abs/1911.00215} {\path{arXiv:1911.00215}},
  \href {https://doi.org/10.1103/PhysRevD.101.014001}
  {\path{doi:10.1103/PhysRevD.101.014001}}.

\bibitem{Tan:2020ldi}
Y.~Tan, W.~Lu, J.~Ping, {Systematics of $QQ\bar q\bar q$ in a chiral
  constituent quark model}, Eur. Phys. J. Plus 135~(9) (2020) 716.
\newblock \href {http://arxiv.org/abs/2004.02106} {\path{arXiv:2004.02106}},
  \href {https://doi.org/10.1140/epjp/s13360-020-00741-w}
  {\path{doi:10.1140/epjp/s13360-020-00741-w}}.

\bibitem{Chen:2013aba}
W.~Chen, T.~G. Steele, S.-L. Zhu, {Exotic open-flavor $bc\bar q\bar q$, $bc\bar
  s\bar s$ and $qc\bar q\bar b$, $sc\bar s\bar b$ tetraquark states}, Phys.
  Rev. D 89~(5) (2014) 054037.
\newblock \href {http://arxiv.org/abs/1310.8337} {\path{arXiv:1310.8337}},
  \href {https://doi.org/10.1103/PhysRevD.89.054037}
  {\path{doi:10.1103/PhysRevD.89.054037}}.

\bibitem{Agaev:2018khe}
S.~S. Agaev, K.~Azizi, B.~Barsbay, H.~Sundu, {Weak decays of the axial-vector
  tetraquark $T^-_{bb;\bar u\bar d}$}, Phys. Rev. D 99~(3) (2019) 033002.
\newblock \href {http://arxiv.org/abs/1809.07791} {\path{arXiv:1809.07791}},
  \href {https://doi.org/10.1103/PhysRevD.99.033002}
  {\path{doi:10.1103/PhysRevD.99.033002}}.

\bibitem{Francis:2018jyb}
A.~Francis, R.~J. Hudspith, R.~Lewis, K.~Maltman, {Evidence for charm-bottom
  tetraquarks and the mass dependence of heavy-light tetraquark states from
  lattice QCD}, Phys. Rev. D 99~(5) (2019) 054505.
\newblock \href {http://arxiv.org/abs/1810.10550} {\path{arXiv:1810.10550}},
  \href {https://doi.org/10.1103/PhysRevD.99.054505}
  {\path{doi:10.1103/PhysRevD.99.054505}}.

\bibitem{Padmanath:2023rdu}
M.~Padmanath, A.~Radhakrishnan, N.~Mathur, {Bound Isoscalar Axial-Vector
  $bc\bar u\bar d$ Tetraquark $T_{bc}$ from Lattice QCD Using Two-Meson and
  Diquark-Antidiquark Variational Basis}, Phys. Rev. Lett. 132~(20) (2024)
  201902.
\newblock \href {http://arxiv.org/abs/2307.14128} {\path{arXiv:2307.14128}},
  \href {https://doi.org/10.1103/PhysRevLett.132.201902}
  {\path{doi:10.1103/PhysRevLett.132.201902}}.

\bibitem{Alexandrou:2023cqg}
C.~Alexandrou, J.~Finkenrath, T.~Leontiou, S.~Meinel, M.~Pflaumer, M.~Wagner,
  {Shallow Bound States and Hints for Broad Resonances with Quark Content
  $b\bar c\bar ud$ in $B$-$\bar D$ and $B^*$-$\bar D$ Scattering from Lattice
  QCD}, Phys. Rev. Lett. 132~(15) (2024) 151902.
\newblock \href {http://arxiv.org/abs/2312.02925} {\path{arXiv:2312.02925}},
  \href {https://doi.org/10.1103/PhysRevLett.132.151902}
  {\path{doi:10.1103/PhysRevLett.132.151902}}.

\bibitem{Born:1927boa}
M.~Born, R.~Oppenheimer, {Zur Quantentheorie der Molekeln}, Annalen Phys.
  389~(20) (1927) 457--484.
\newblock \href {https://doi.org/10.1002/andp.19273892002}
  {\path{doi:10.1002/andp.19273892002}}.

\bibitem{Braaten:2014qka}
E.~Braaten, C.~Langmack, D.~H. Smith, {Born-Oppenheimer Approximation for the
  XYZ Mesons}, Phys. Rev. D 90~(1) (2014) 014044.
\newblock \href {http://arxiv.org/abs/1402.0438} {\path{arXiv:1402.0438}},
  \href {https://doi.org/10.1103/PhysRevD.90.014044}
  {\path{doi:10.1103/PhysRevD.90.014044}}.

\bibitem{Juge:1999ie}
K.~J. Juge, J.~Kuti, C.~J. Morningstar, {Ab initio study of hybrid $\bar b g b$
  mesons}, Phys. Rev. Lett. 82 (1999) 4400--4403.
\newblock \href {http://arxiv.org/abs/hep-ph/9902336}
  {\path{arXiv:hep-ph/9902336}}, \href
  {https://doi.org/10.1103/PhysRevLett.82.4400}
  {\path{doi:10.1103/PhysRevLett.82.4400}}.

\bibitem{Brambilla:2017uyf}
N.~Brambilla, G.~Krein, J.~Tarr{\'u}s~Castell{\`a}, A.~Vairo, {Born-Oppenheimer
  approximation in an effective field theory language}, Phys. Rev. D 97~(1)
  (2018) 016016.
\newblock \href {http://arxiv.org/abs/1707.09647} {\path{arXiv:1707.09647}},
  \href {https://doi.org/10.1103/PhysRevD.97.016016}
  {\path{doi:10.1103/PhysRevD.97.016016}}.

\bibitem{Berwein:2024ztx}
M.~Berwein, N.~Brambilla, A.~Mohapatra, A.~Vairo, {Hybrids, tetraquarks,
  pentaquarks, doubly heavy baryons, and quarkonia in Born-Oppenheimer
  effective theory}, Phys. Rev. D 110~(9) (2024) 094040.
\newblock \href {http://arxiv.org/abs/2408.04719} {\path{arXiv:2408.04719}},
  \href {https://doi.org/10.1103/PhysRevD.110.094040}
  {\path{doi:10.1103/PhysRevD.110.094040}}.

\bibitem{Brambilla:2024imu}
N.~Brambilla, A.~Mohapatra, T.~Scirpa, A.~Vairo, {Nature of $\chi_{c1}(3872)$
  and $T_{cc}^+(3875)$}, Phys. Rev. Lett. 135~(13) (2025) 131902.
\newblock \href {http://arxiv.org/abs/2411.14306} {\path{arXiv:2411.14306}},
  \href {https://doi.org/10.1103/pdy7-hvg7} {\path{doi:10.1103/pdy7-hvg7}}.

\bibitem{Brambilla:2025xma}
N.~Brambilla, A.~Mohapatra, A.~Vairo, {Unraveling pentaquarks with the
  Born-Oppenheimer effective theory}, Phys. Rev. D 112~(11) (2025) 114037.
\newblock \href {http://arxiv.org/abs/2508.13050} {\path{arXiv:2508.13050}},
  \href {https://doi.org/10.1103/5z3t-rq5f} {\path{doi:10.1103/5z3t-rq5f}}.

\bibitem{Allaman:2024vwn}
H.~Allaman, M.~Ekhterachian, F.~Nardi, R.~Rattazzi, S.~Stelzl, {Tetraquarks at
  large $M$ and large $N$}, JHEP 11 (2024) 034.
\newblock \href {http://arxiv.org/abs/2407.18298} {\path{arXiv:2407.18298}},
  \href {https://doi.org/10.1007/JHEP11(2024)034}
  {\path{doi:10.1007/JHEP11(2024)034}}.

\bibitem{Maiani:2022qze}
L.~Maiani, A.~Pilloni, A.~D. Polosa, V.~Riquer, {Doubly heavy tetraquarks in
  the Born-Oppenheimer approximation}, Phys. Lett. B 836 (2023) 137624.
\newblock \href {http://arxiv.org/abs/2208.02730} {\path{arXiv:2208.02730}},
  \href {https://doi.org/10.1016/j.physletb.2022.137624}
  {\path{doi:10.1016/j.physletb.2022.137624}}.

\bibitem{Juge:2002br}
K.~J. Juge, J.~Kuti, C.~Morningstar, {Fine structure of the QCD string
  spectrum}, Phys. Rev. Lett. 90 (2003) 161601.
\newblock \href {http://arxiv.org/abs/hep-lat/0207004}
  {\path{arXiv:hep-lat/0207004}}, \href
  {https://doi.org/10.1103/PhysRevLett.90.161601}
  {\path{doi:10.1103/PhysRevLett.90.161601}}.

\bibitem{Mutuk:2025hql}
H.~Mutuk, {Exotic $T_{c\bar s0}^a(2900)^0$ and $T_{c\bar s0}^a(2900)^{++}$
  states in the Born-Oppenheimer approximation}, Phys. Rev. D 113~(3) (2026)
  034020.
\newblock \href {http://arxiv.org/abs/2512.01028} {\path{arXiv:2512.01028}},
  \href {https://doi.org/10.1103/3z79-67zf} {\path{doi:10.1103/3z79-67zf}}.

\bibitem{Anselmino:1992vg}
M.~Anselmino, E.~Predazzi, S.~Ekelin, S.~Fredriksson, D.~B. Lichtenberg,
  {Diquarks}, Rev. Mod. Phys. 65 (1993) 1199--1234.
\newblock \href {https://doi.org/10.1103/RevModPhys.65.1199}
  {\path{doi:10.1103/RevModPhys.65.1199}}.

\bibitem{Barabanov:2020jvn}
M.~Y. Barabanov, et~al., {Diquark correlations in hadron physics: Origin,
  impact and evidence}, Prog. Part. Nucl. Phys. 116 (2021) 103835.
\newblock \href {http://arxiv.org/abs/2008.07630} {\path{arXiv:2008.07630}},
  \href {https://doi.org/10.1016/j.ppnp.2020.103835}
  {\path{doi:10.1016/j.ppnp.2020.103835}}.

\bibitem{Lodha:2024fbs}
C.~Lodha, A.~K. Rai, {Investigation of Mass and Decay Characteristics of the
  All-light Tetraquark}, Few Body Syst. 65~(4) (2024) 99.
\newblock \href {http://arxiv.org/abs/2410.14246} {\path{arXiv:2410.14246}},
  \href {https://doi.org/10.1007/s00601-024-01968-4}
  {\path{doi:10.1007/s00601-024-01968-4}}.

\bibitem{Lodha:2025ijm}
C.~Lodha, M.~Parmar, A.~K. Rai, {Exploring charm bound states: Mass spectra and
  decay dynamics of $D$ mesons and $cq\bar q\bar q$ tetraquarks}, Int. J. Mod.
  Phys. A 40~(28) (2025) 2550125.
\newblock \href {http://arxiv.org/abs/2505.22393} {\path{arXiv:2505.22393}}.

\bibitem{Mohan:2026blk}
B.~Mohan, R.~Dhir, {A baryon-calibrated unified quark-diquark effective mass
  formalism for heavy multiquarks} (2026).
\newblock \href {http://arxiv.org/abs/2603.01037} {\path{arXiv:2603.01037}}.

\bibitem{Bali:2000un}
G.~S. Bali, {Casimir scaling of SU(3) static potentials}, Phys. Rev. D 62
  (2000) 114503.
\newblock \href {http://arxiv.org/abs/hep-lat/0006022}
  {\path{arXiv:hep-lat/0006022}}, \href
  {https://doi.org/10.1103/PhysRevD.62.114503}
  {\path{doi:10.1103/PhysRevD.62.114503}}.

\bibitem{Brodsky:2014xia}
S.~J. Brodsky, D.~S. Hwang, R.~F. Lebed, {Dynamical Picture for the Formation
  and Decay of the Exotic XYZ Mesons}, Phys. Rev. Lett. 113~(11) (2014) 112001.
\newblock \href {http://arxiv.org/abs/1406.7281} {\path{arXiv:1406.7281}},
  \href {https://doi.org/10.1103/PhysRevLett.113.112001}
  {\path{doi:10.1103/PhysRevLett.113.112001}}.

\bibitem{Lebed:2017min}
R.~F. Lebed, {Spectroscopy of Exotic Hadrons Formed from Dynamical Diquarks},
  Phys. Rev. D 96~(11) (2017) 116003.
\newblock \href {http://arxiv.org/abs/1709.06097} {\path{arXiv:1709.06097}},
  \href {https://doi.org/10.1103/PhysRevD.96.116003}
  {\path{doi:10.1103/PhysRevD.96.116003}}.

\bibitem{Giron:2019bcs}
J.~F. Giron, R.~F. Lebed, C.~T. Peterson, {The Dynamical Diquark Model: First
  Numerical Results}, JHEP 05 (2019) 061.
\newblock \href {http://arxiv.org/abs/1903.04551} {\path{arXiv:1903.04551}},
  \href {https://doi.org/10.1007/JHEP05(2019)061}
  {\path{doi:10.1007/JHEP05(2019)061}}.

\bibitem{Giron:2019cfc}
J.~F. Giron, R.~F. Lebed, C.~T. Peterson, {The Dynamical Diquark Model: Fine
  Structure and Isospin}, JHEP 01 (2020) 124.
\newblock \href {http://arxiv.org/abs/1907.08546} {\path{arXiv:1907.08546}},
  \href {https://doi.org/10.1007/JHEP01(2020)124}
  {\path{doi:10.1007/JHEP01(2020)124}}.

\bibitem{Giron:2020qpb}
J.~F. Giron, R.~F. Lebed, {Spectrum of the hidden-bottom and the
  hidden-charm-strange exotics in the dynamical diquark model}, Phys. Rev. D
  102~(1) (2020) 014036.
\newblock \href {http://arxiv.org/abs/2005.07100} {\path{arXiv:2005.07100}},
  \href {https://doi.org/10.1103/PhysRevD.102.014036}
  {\path{doi:10.1103/PhysRevD.102.014036}}.

\bibitem{Giron:2020wpx}
J.~F. Giron, R.~F. Lebed, {Simple spectrum of $c\bar c c\bar c$ states in the
  dynamical diquark model}, Phys. Rev. D 102~(7) (2020) 074003.
\newblock \href {http://arxiv.org/abs/2008.01631} {\path{arXiv:2008.01631}},
  \href {https://doi.org/10.1103/PhysRevD.102.074003}
  {\path{doi:10.1103/PhysRevD.102.074003}}.

\bibitem{Wang:2011ab}
Z.-G. Wang, {Analysis of the light-flavor scalar and axial-vector diquark
  states with QCD sum rules}, Commun. Theor. Phys. 59 (2013) 451--456.
\newblock \href {http://arxiv.org/abs/1112.5910} {\path{arXiv:1112.5910}},
  \href {https://doi.org/10.1088/0253-6102/59/4/11}
  {\path{doi:10.1088/0253-6102/59/4/11}}.

\bibitem{Feng:2023txx}
X.~Feng, J.-K. Chen, J.-Q. Xie, {Regge trajectories for the doubly heavy
  diquarks}, Phys. Rev. D 108~(3) (2023) 034022.
\newblock \href {http://arxiv.org/abs/2305.15705} {\path{arXiv:2305.15705}},
  \href {https://doi.org/10.1103/PhysRevD.108.034022}
  {\path{doi:10.1103/PhysRevD.108.034022}}.

\bibitem{Maiani:2004vq}
L.~Maiani, F.~Piccinini, A.~D. Polosa, V.~Riquer, {Diquark-antidiquarks with
  hidden or open charm and the nature of X(3872)}, Phys. Rev. D 71 (2005)
  014028.
\newblock \href {http://arxiv.org/abs/hep-ph/0412098}
  {\path{arXiv:hep-ph/0412098}}, \href
  {https://doi.org/10.1103/PhysRevD.71.014028}
  {\path{doi:10.1103/PhysRevD.71.014028}}.

\bibitem{Esau:2019hqw}
S.~Esau, A.~Palameta, R.~T. Kleiv, D.~Harnett, T.~G. Steele, {Axial Vector $cc$
  and $bb$ Diquark Masses from QCD Laplace Sum-Rules}, Phys. Rev. D 100 (2019)
  074025.
\newblock \href {http://arxiv.org/abs/1905.12803} {\path{arXiv:1905.12803}},
  \href {https://doi.org/10.1103/PhysRevD.100.074025}
  {\path{doi:10.1103/PhysRevD.100.074025}}.

\bibitem{Dowdall:2012ab}
R.~J. Dowdall, C.~T.~H. Davies, T.~C. Hammant, R.~R. Horgan, {Precise
  heavy-light meson masses and hyperfine splittings from lattice QCD including
  charm quarks in the sea}, Phys. Rev. D 86 (2012) 094510.
\newblock \href {http://arxiv.org/abs/1207.5149} {\path{arXiv:1207.5149}},
  \href {https://doi.org/10.1103/PhysRevD.86.094510}
  {\path{doi:10.1103/PhysRevD.86.094510}}.

\bibitem{ParticleDataGroup:2024cfk}
S.~Navas, et~al., {Review of particle physics}, Phys. Rev. D 110~(3) (2024)
  030001.
\newblock \href {https://doi.org/10.1103/PhysRevD.110.030001}
  {\path{doi:10.1103/PhysRevD.110.030001}}.

\bibitem{Karliner:2017qjm}
M.~Karliner, J.~L. Rosner, {Discovery of doubly-charmed $\Xi_{cc}$ baryon
  implies a stable $(bb\bar u\bar d)$ tetraquark}, Phys. Rev. Lett. 119~(20)
  (2017) 202001.
\newblock \href {http://arxiv.org/abs/1707.07666} {\path{arXiv:1707.07666}},
  \href {https://doi.org/10.1103/PhysRevLett.119.202001}
  {\path{doi:10.1103/PhysRevLett.119.202001}}.

\bibitem{Guo:2021yws}
T.~Guo, J.~Li, J.~Zhao, L.~He, {Mass spectra of doubly heavy tetraquarks in an
  improved chromomagnetic interaction model}, Phys. Rev. D 105~(1) (2022)
  014021.
\newblock \href {http://arxiv.org/abs/2108.10462} {\path{arXiv:2108.10462}},
  \href {https://doi.org/10.1103/PhysRevD.105.014021}
  {\path{doi:10.1103/PhysRevD.105.014021}}.

\end{thebibliography}

\end{document}